\documentclass[a4paper,11pt]{article}

\usepackage{jcappub}
\usepackage{color}
\allowdisplaybreaks
\usepackage{graphicx}
\usepackage{epstopdf}

\newcommand{\be}{\begin{equation}}
\newcommand{\ee}{\end{equation}}
\newcommand{\bea}{\begin{eqnarray}}
\newcommand{\eea}{\end{eqnarray}}
\newcommand{\nn}{\nonumber}
\newcommand{\bdm}{\begin{displaymath}}
\newcommand{\edm}{\end{displaymath}}

\title{Massive neutrinos in 
\\nonlinear large scale structure:\\ 
A consistent perturbation theory}

\author[a,b]{Michele Levi}
\author[c,d]{and Zvonimir Vlah}

\affiliation[a]{Sorbonne Universit\'es, 
Universit\'e Pierre et Marie Curie--Paris 6, CNRS, UMR 
7095,\\
Institut d'astrophysique de Paris, 
98bis boulevard Arago, 75014 Paris, France} 
\affiliation[b]{Sorbonne Universit\'es, Institut Lagrange 
de Paris,\\ 
	98bis boulevard Arago, 75014 Paris, France} 
\affiliation[c]{Stanford Institute for Theoretical 
Physics, 
Stanford University, Stanford, CA 94305, USA}
\affiliation[d]{Kavli Institute for Particle Astrophysics 
and Cosmology, 
Stanford University and SLAC, \\ Menlo Park, CA 94025, 
USA}

\emailAdd{michele.levi@upmc.fr}
\emailAdd{zvlah@stanford.edu}

\abstract{A consistent formulation to incorporate massive neutrinos in 
the perturbation theory of the effective CDM+baryons fluid is 
introduced. In this formulation all linear k dependence in the growth 
functions of CDM+baryons perturbations, as well as all consequent 
additional mode coupling at higher orders, are taken into account to any 
desirable accuracy. Our formulation regards the neutrino fraction, which 
is constant in time after the non-relativistic transition of neutrinos, 
and much smaller than unity, as the coupling constant of the theory. 
Then the ``bare'' perturbations are those in the massless neutrino case 
when the neutrino fraction vanishes, and we consider the backreaction 
corrections due to the gravitational coupling of neutrinos. We derive 
the general equations for the ``bare'' perturbations, and backrecation corrections. 
Then, by employing exact time evolution with the proper analytic Green's 
function we explicitly derive the leading backreaction effect, and find 
precise agreement at the linear level. We proceed to compute the second 
order beackreaction correction, and derive the leading order matter
bispectrum in the presence of massive neutrinos, suggesting the squeezed 
limit of the matter bispectrum as a sensitive probe of neutrino mass.
Notably, the generic neutrino fraction formulation in this work may be 
similarly applied for the consistent inclusion of massive neutrinos 
within any perturbative approach.} 


\begin{document}


\maketitle

\flushbottom

\section{Introduction}
\label{intro}

Future large scale structure surveys, such as the Euclid 
mission \cite{EC,Laureijs:2011gra,  Amendola:2012ys}, or 
ground-based LSST \cite{LSST}, are expected to be the next 
leading probe of cosmological information. Beyond addressing 
the puzzling cosmological questions of the accelerated 
expansion of the Universe, i.e.~the nature of Dark Energy, or 
the primordial fluctuations in the early Universe, they aim to 
explore questions that extend to other fields of fundamental 
physics, such as the composition of Dark Matter and neutrinos 
beyond the Standard Model of particle physics, or tests of 
alternative theories of Gravity. The goal of these future 
surveys is to reach a 1\% observational accuracy, thus
requiring similar high precision theoretical modeling of the 
relevant observables.

In particular, cosmological data may enable us to fix the 
absolute neutrino mass scale, and the number of neutrino 
species. Whereas the total neutrino mass $M_\nu \equiv \sum_i 
m_{\nu,i}$, with $m_{\nu,i}$ the mass of each neutrino species, 
has a lower bound of $M_\nu>0.06$ eV coming from neutrino 
oscillation data \cite{GonzalezGarcia:2012sz}, it is the 
combination of cosmological data from the cosmic 
microwave background (CMB) and large scale structure (LSS) 
observations, which currently provides the stringent upper 
bound to this quantity at $M_\nu<0.14$ eV \cite{Costanzi:2014tna, Ade:2013zuv}. 
In fact, 
cosmological structure formation is quite sensitive to the 
neutrino mass, and even the smallest possible total neutrino 
mass in the currently bound range has an impact of at least 5\% 
on the total matter power spectrum across nonlinear scales, 
where the neutrino effect is maximal. Hence, in turn, in order 
to extract more information from galaxy surveys to also fix all 
cosmological parameters with high precision, it is essential to 
improve our understanding of linear and nonlinear structure 
formation, allowing to increase the volume of related Fourier 
space in realistic massive neutrino cosmologies.

Specifically nonlinear higher order corrections to n-point 
functions, in particular the total matter power spectrum in the 
presence of massive neutrinos, has been studied both 
analytically using perturbation theories \cite{Bernardeau:2001qr}, e.g.~in 
\cite{Saito:2008bp,Wong:2008ws,Lesgourgues:2009am,Saito:2009ah,Upadhye:2013ndm,Blas:2014hya}, 
and with N-body simulations, e.g.~in the more recent 
\cite{Bird:2011rb,Castorina:2015bma}. The leading order bispectrum 
has also been approached analytically in \cite{Fuhrer:2014zka}. 
All previous analytical work treat cold dark 
matter (CDM) combined with baryons as an ideal pressureless 
fluid, which reproduces the departure from linear theory in a 
very limited range of scales, even in the case of massless 
neutrinos, roughly $k\lesssim0.1 h\,Mpc^{-1}$ at $z=0$. The 
works \cite{Saito:2008bp,Saito:2009ah,Wong:2008ws,Fuhrer:2014zka,Dupuy:2013jaa},
which are based on standard perturbation theory (SPT) 
\cite{Bernardeau:2001qr} consider an Einstein de Sitter (EdS) 
Universe as the baseline cosmology, rather than handle the exact 
$\Lambda$CDM time evolution. Other works 
\cite{Lesgourgues:2009am,Upadhye:2013ndm,Blas:2014hya} rely 
on the time renormalization group (TRG) flow approach , where 
the perturbative (loop) order is not well-defined. Previous works 
have ignored the k dependence of linear growth functions of the 
CDM+baryons perturbations in the presence of massive neutrinos, and the 
additional consequent mode coupling at higher orders, or considered them 
in an incomplete manner. 
N-body simulations have obtained the total matter power spectrum at 
low redshifts up to the present $z=0$, and reach large 
$k$ modes in the fully non-linear regime up to $k\sim 10 \,h\, 
Mpc^{-1}$. It should be stressed though that even if N-body 
simulations succeed in modeling nonlinear observables to high 
precision, they must be grounded in theory, which extends to 
the mildly nonlinear regime.

Most previous analytical and numerical works treated neutrinos as 
a linear perturbation, which acts as an external gravitational 
source. All past works, both in analytical methods, and N-body 
simulations, concluded that on the relevant scales, and 
considering the constraints on the total neutrino mass, 
the total nonlinear matter power spectrum and leading order 
bispectrum in massive neutrino cosmologies can be 
described at the 1\% level by accounting for the nonlinear 
evolution of CDM perturbations alone, while adopting the linear 
approximation for the neutrino component.
In \cite{Blas:2014hya} a fluid description was assumed for the 
neutrino component in order to evaluate its nonlinear 
evolution. However, on length scales smaller than the characteristic 
free streaming scale, $k_{FS}$, the 
fluid description formally leads to acoustic oscillations in 
the neutrino density contrast, which renders the fluid approach 
a poor description of the clustering behavior of the 
free-streaming dark matter (DM). Indeed, in  \cite{Fuhrer:2014zka}, 
which also set out to evaluate the nonlinear evolution of the 
neutrino component and to examine the validity of the fluid 
approximation on the transitional length scales $k\sim k_{FS}$, 
especially at higher perturbative orders, an enhanced breakdown 
of the fluid approximation at the nonlinear level was found for 
the relevant neutrino mass range, as all relevant 
observable $k$ modes are greater than $k_{FS}$. Finally, we 
also note that some of the analytical methods \cite{Dupuy:2013jaa, Fuhrer:2014zka}, as well as 
N-body simulations, are computationally intensive, and hence 
practically not favorable.

In this work we introduce a consistent formulation for the 
inclusion of massive neutrinos in the evolution of
matter perturbations, such that all linear k dependence
of the CDM+baryons perturbations, as well as all additional 
mode coupling at higher orders, are taken into account to any 
desirable accuracy. This is done using the fact that the linear 
neutrino component, can be represented to any desirable 
accuracy as a sum of separable functions of $k$, the scale 
factor $a$, and a generic $f_{\nu}$ dependence, with the required 
asymptotic behavior.
Our formulation, is based on the fact that after the 
non-relativistic (NR) transition of the neutrinos, the neutrino 
fraction, $f_\nu$, is constant in time, and very small, 
particularly so in light of the current constrained range of total 
neutrino mass in eq.~\eqref{numrange}. Therefore $f_\nu$ is 
ideal for use as the coupling constant of the theory including 
massive neutrinos, and our formulation is made in terms of 
a generic $f_\nu$. This is obviously advantageous in order 
to explore the possible range of this parameter. Then the ``bare'' 
perturbations are those in the massless neutrino case when the 
neutrino fraction vanishes, and we consider the backreaction 
corrections due to the gravitational coupling of massive neutrinos.
Further, our formulation resides within the general perturbative 
view of the CDM and baryons as an effective 
fluid \cite{Baumann:2010tm,Carrasco:2012cv}, that is within a 
well-defined perturbative theory, reaching a $1\%$ accuracy for 
small redshifts and larger k modes, i.e.~beyond 
$k\simeq0.1\,h\,Mpc^{-1}$, in the mildly non-linear regime. We stress 
though that the ingredients presented here are generic, and can equally 
be applied in other perturbative approaches.
We employ here an exact time evolution, using the proper 
explicit analytic Green's function similar to the $\Lambda$CDM 
baseline cosmology, common to all $f_\nu$ cosmologies after the 
NR transition of neutrinos, rather than resorting to uncontrolled 
EdS-like approximations. This is done while maintaining computational 
efficiency for practical use.

The outline of the paper is as follows. In section \ref{NC}
we begin with reviewing basic neutrino physics, focusing in 
section \ref{nulss} on their impact on structure formation. In 
section \ref{ptnu}, we start by highlighting the main 
ingredients in our formulation, which focuses on the exact evolution of 
the CDM+baryons perturbations. In section \ref{eqbr} 
we introduce the ``bare'' and backreaction perturbations,  
derive their general equations, and show the required exact 
time evolution, using the proper explicit analytic Green's function. 
Then in section \ref{lobr} we explicitly derive the leading 
backreaction effect, and find precise agreement with the linear 
effect, and the linear total matter power spectrum. In section 
\ref{lonubis} we proceed to compute the second order 
backreaction correction, and derive the leading order matter 
bispectrum in the presence of massive neutrinos, where we 
explore the suppression effect in the shape dependence of the 
bispectrum. Further, in section \ref{exactnlnu} we discuss the relevance 
of an exact evaluation of the nonlinear neutrino perturbations for high 
precision nonlinear LSS in the presence of massive neutrinos. In 
section \ref{theendmyfriend} we summarize our 
main conclusions. Finally, in appendix \ref{useapxs} we provide 
useful numerical approximations for possible efficient accurate 
numerical evaluations in higher orders, and in appendix \ref{realiz} 
we provide the details of the realizations implemented in this work.

\section{Neutrino Cosmology} 
\label{NC}

Let us review basic neutrino physics in cosmology, in 
particular their role in structure formation 
\cite{Lesgourgues2013,Lesgourgues:2006nd,Shoji:2010hm} 

Massive neutrinos are considered hot dark matter since 
they decouple as relativistic particles in the early 
Universe, just before the onset of Big Bang 
Nucleosynthesis. The mass density of the massive 
neutrinos after they have 
become non-relativistic is given by
\be
\Omega_\nu h^2 = \frac{M_\nu}{93.14\,eV}.
\ee 
The current bounds on the sum of neutrino masses, which  
are not included in the Standard Model of particle 
physics, are given by
\be \label{numrange}
0.06\, eV \lesssim M_\nu\equiv\Sigma\, m_\nu \lesssim 
0.14\,eV. 
\ee
The lower bound comes from solar and atmospheric neutrino 
oscillation experiments \cite{GonzalezGarcia:2012sz}, as 
discovered in 1998, whereas the most stringent upper 
bounds come from Cosmology, provided by constraints from 
the combination of recent CMB and LSS data 
\cite{Ade:2013zuv,Costanzi:2014tna}. 

The ratio of the neutrino density to the total matter 
density, namely the neutrino fraction $f_{\nu}$, is given 
by
\be
f_\nu\equiv\frac{\Omega_\nu}{\Omega_m}\lesssim0.01,
\ee
and is constant once the neutrinos have become 
non-relativistic (NR). This fact, together with the smallness 
of $f_\nu$, play a key role in our ability to formulate a 
perturbation theory for the inclusion of  
massive neutrinos in structure formation, 
using $f_\nu$ as the small coupling constant of the 
theory, as we illustrate in section \ref{ptnu}. 

The massive neutrinos have a large velocity dispersion, 
$\sigma_\nu$, following their frozen Fermi-Dirac 
distribution since their decoupling. The non-relativistic 
transition of the neutrinos occurs when the mean neutrino 
energy becomes smaller than the neutrino mass, i.e.
\begin{align}
\langle E \rangle=\frac{\int d^3p \, p\, 
[(exp(p/T_{\nu}(z))+1]^{-1}}
{\int d^3p\, [(exp(p/T_{\nu}(z))+1]^{-1}}
\simeq 3.15 T_{\nu,0}(1+z)\leq m_\nu,
\end{align}
hence at a redshift given by  
\be
1+z_{nr}\simeq\frac{m_{\nu}}{5.28\times10^{-4} \,eV}.
\ee	 
After the non-relativistic transition the neutrino 
thermal velocity decays with time like 
\be
\sigma_\nu\simeq 
150(1+z)\left[\frac{1\,eV}{m_\nu}\right]km\,s^{-1},
\ee
which is just the analogue of the cosmological redshift 
for a massive particle, and  we note that the 
non-relativistic transition occurs when $\sigma_\nu 
\simeq 0.3\,c$, where $c$ is the speed of light.

\subsection{Massive neutrinos and structure formation}
\label{nulss}

If we write a fluid equation for the neutrino 
component, we get
\be
\dot{\theta}(\vec{k},\tau)+\mathcal{H}(\tau)\theta(\vec{k}
,\tau)
+\left(\frac{3}{2}\mathcal{H}^2(\tau)-k^2c_s^2(\tau)\right
)
\delta(\vec{k},\tau)=0,
\ee
where $\delta$ and $\theta$ denote the usual density 
contrast and velocity divergence, respectively.
By analogy with the Jeans length, the neutrino velocity 
dispersion introduces a further time dependent dynamical 
scale into the problem, usually referred as the 
``free-streaming scale'', $\lambda_{FS}\equiv 2\pi 
ak_{FS}^{-1}$, which corresponds to the free-streaming 
wavenumber given by 
\be\label{kfs}
k_{FS}(z)\equiv\sqrt{\frac{3
	}{2}}\frac{\mathcal{H}(z)}{c_s(z)}\simeq\sqrt{\frac{3
		}{2}}\frac{\mathcal{H}(z)}{\sigma_\nu(z)}.
\ee
This is the scale, below which collisionless particles 
cannot remain confined in gravitational potential wells, 
because of their velocity dispersion. 

It should be noted that another integrated quantity is 
useful to describe the scale, above which neutrino 
free-streaming can be completely ignored. This is defined 
like any other comoving horizon scale:
\be
d_{FS}(\eta)\equiv 
a(\eta)\int_{\eta_{dec}}^{\eta}\sigma_{\nu}(\eta)d\eta, 
\ee
which we refer to as the ``free-streaming horizon''. This 
gives the average distance traveled by neutrinos between 
the early universe and a given time. As long as the 
neutrinos are relativistic it is easy to see that the two 
scales are similar as the comoving free-streaming scale 
$k_{FS}^{-1}$ grows closely with the comoving Hubble 
scale $\mathcal{H}^{-1}$, until the non-relativistic 
transition of the neutrino.

However once the neutrino becomes non-relativistic the 
comoving free-streaming scale $k_{FS}^{-1}$ starts 
decreasing. Thus there is a maximal comoving 
free-streaming scale, corresponding to the wavenumber 
denoted by $k_{nr}$, which is set by the minimal value of 
$k_{FS}$ at the time of the non-relativistic transition. 
This scale is approximated by
\be 
k_{nr}\equiv 
k_{FS}(z_{nr})\simeq0.018\sqrt{\Omega_m\left[\frac{m_\nu}{
1\,eV}\right]}
\,h\,Mpc^{-1}\lesssim 5\times10^{-3}\,h\,Mpc^{-1}.
\ee
Hence, modes with $k \lesssim k_{nr}$ are not affected by 
free-streaming, and evolve like in a pure $\Lambda$CDM 
cosmology. Yet, all relevant observable $k$ modes are 
greater than $k_{nr}$. 

\begin{figure}[t]
\begin{centering}
\includegraphics[scale=0.8]{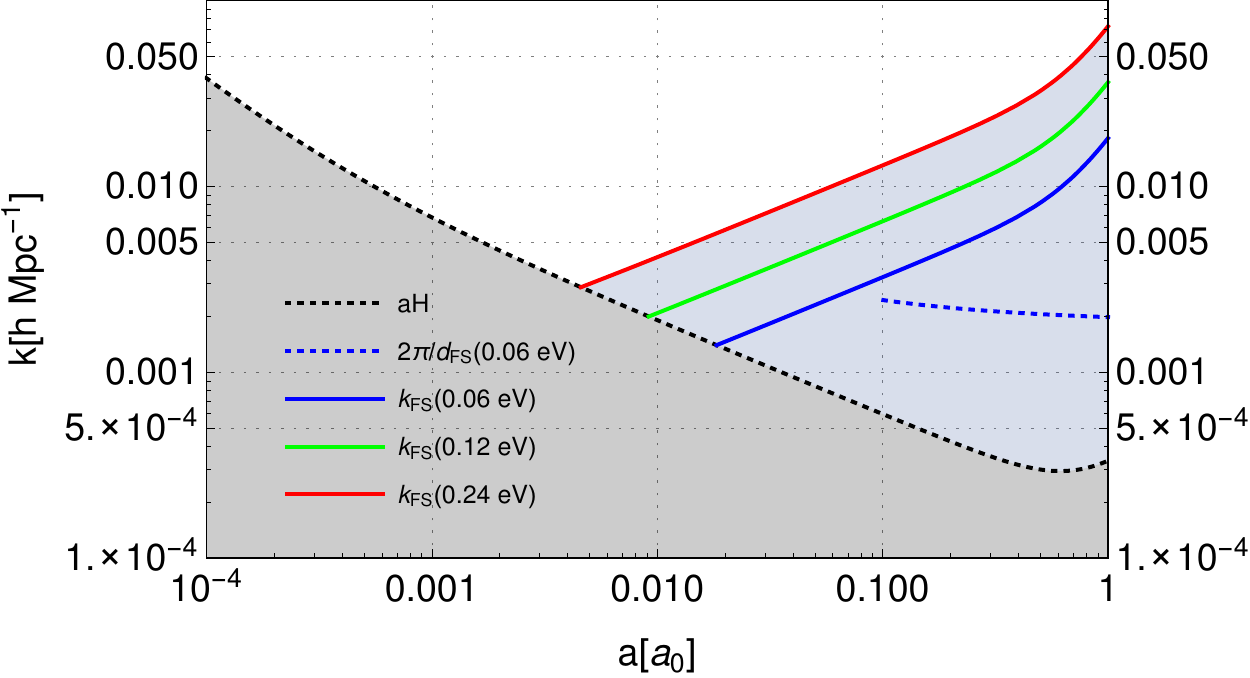}
\caption{The transition to a non-relativistic free 
streaming scale of a massive neutrino as a 
function of the scale factor, a, for the total neutrino 
mass values: $M_{\nu}=0.06, 0.12, 0.24$ eV, and 3 
degenerate massive species. The black dotted line is the comoving 
horizon scale. The blue dotted line shows the free streaming horizon 
scale for $M_{\nu}=0.06$ eV after the NR transition at late times. 
The colored lines are approximations to the free streaming 
scale, $k_{FS}$, in the NR limit, which hold only 
\textit{after} the NR transition time, 
roughly when the lines intersect with the comoving 
horizon scale. Lighter neutrinos become NR 
at later times, and in the NR limit the free streaming 
scale is just proportional to the neutrino mass, 
$k_{FS}\propto m_{\nu}$. The neutrino density 
fluctuations start to grow only when their modes enter 
the bluish region of the plot, e.g.~for the $M_{\nu}=0.24$ 
eV case. Note that the relevant characteristic scale is in fact 
that, which corresponds the free streaming horizon. This scale 
still decreases with time, but remains close to $k_{nr}$, the 
minimal $k_{FS}$, after the NR transition. Hence, the characteristic 
free streaming scale is taken as $k_{nr}\simeq5\times10^{-3}$, notably 
below current observable scales.}\label{khfs}
\end{centering}
\end{figure}

We note that the comoving free-streaming horizon 
$a^{-1}d_{FS}$ increases with time, but that for late 
times it remains very close to the maximal comoving 
free-streaming scale $2\pi k_{nr}^{-1}$. Actually, it is
the comoving free streaming horizon, which is the strict 
scale to consider in order to know above which comoving 
scale free streaming can be completely neglected. 
However, in most of the literature it is $2\pi 
k_{nr}^{-1}$ that is considered, which makes no 
difference in practice for neutrinos becoming 
non-relativistic after the equality of matter and 
radiation. 
\cite{Lesgourgues2013}. For this reason in this work we refer 
to $k_{nr}$ as the characteristic scale.
In figure \ref{khfs} we show the non-relativistic 
transition of the free streaming wavenumber $k_{FS}$, as 
inferred from the comoving Hubble horizon scale, and 
after the non-relativistic transition time from taking 
the non-relativistic limit approximation. 

At wavenumbers $k\gtrsim k_{nr}$ the growth of CDM+baryons is 
suppressed due to the lack of neutrino perturbations, whereas 
at $k\lesssim k_{nr}$ the neutrinos cluster together with 
the CDM and baryons. Thus, the linear growth rate of CDM 
and baryons is scale dependent in the presence of massive 
neutrinos. Moreover, neutrino backreaction effects suppress the 
growth of matter perturbations. During matter domination on 
$k\gg k_{nr}$ the neutrino perturbations do not contribute to 
the gravitational clustering, although they do contribute to 
the homogeneous expansion. The linear suppression of the 
matter power spectrum with massive neutrinos with respect 
to that with massless neutrinos on scales below $k_{nr}$ 
is evaluated by 
\be \label{linsup}
\frac{\Delta P_L^{f_\nu}}{P_L^{f_\nu=0}}\sim -8 f_\nu 
\gtrsim 4\%. 
\ee

\begin{figure}[t]
\begin{centering}
\includegraphics[width=\textwidth]{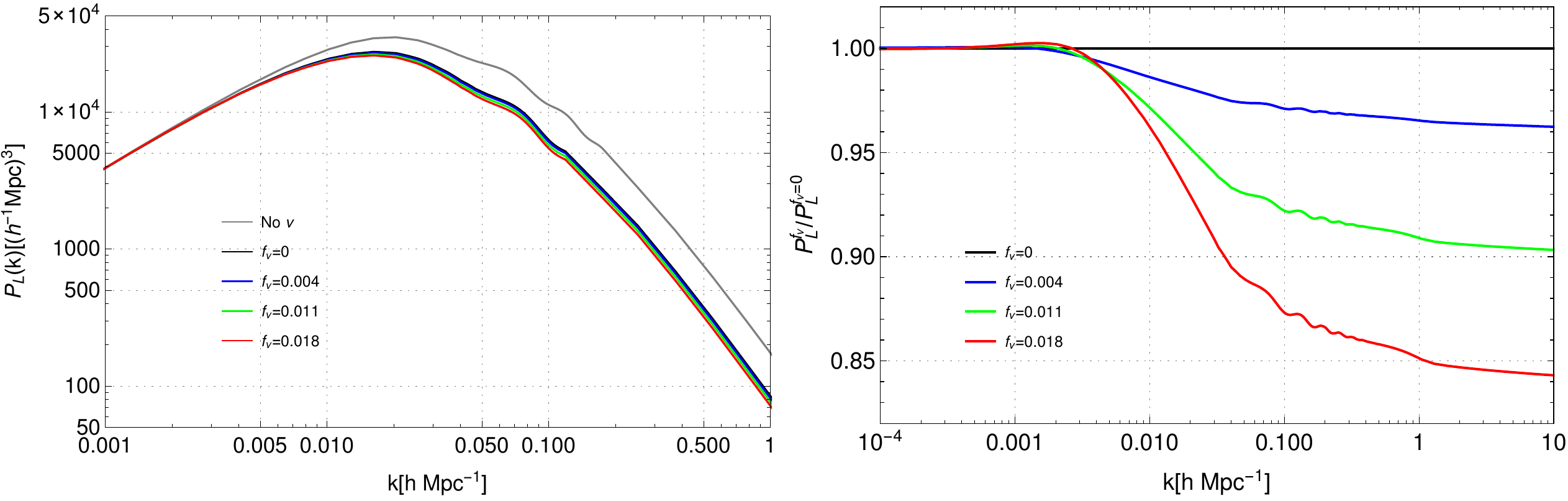}
\caption{\textit{Left:} The linear matter power spectrum in 
the absence of neutrinos, and in the presence of massless 
and massive neutrinos with 3 degenerate species. The 
different solid lines correspond to total neutrino mass 
values $M_{\nu}=0, 0.06, 0.15, 0.24$ eV, parametrized 
by the neutrino fraction $f_{\nu}=0, 0.004, 0.011, 0.018$,
respectively, where the baryon and total matter densities 
are fixed, and the neutrino and CDM densities are varied 
accordingly. \textit{Right:} The suppression of linear 
matter power spectrum with massive neutrinos with respect 
to massless neutrinos. The notation is similar to that on 
the left figure. As expected the suppression grows with the 
total neutrino mass. The minimal wavenumber, where the 
suppression appears is located at 
$k_{nr}\sim10^{-3}-10^{-2}$.}\label{numass}
\end{centering}
\end{figure}

In figure \ref{numass} we show the linear suppression of 
the matter power spectrum for various values of the 
neutrino fraction, corresponding to the relevant mass 
range in eq.~\eqref{numrange}. 
The neutrino effect is maximal beyond the linear regime, 
and appears at very similar scales to BAOs.
From N-body simulations 
\cite{Bird:2011rb,Castorina:2015bma} the non-linear 
suppression of the matter power spectrum is evaluated to
\be \label{nlinsup}
\frac{\Delta P_{NL}^{f_\nu}}{P_{NL}^{f_\nu=0}}\sim -10 
f_\nu \gtrsim 5\%. 
\ee 

In terms of the total matter density contrast, which is 
given by 
\be \label{denmat}
\delta^m=(1-f_\nu)\delta^c+f_\nu\delta^\nu,
\ee
where we denote by the superscript ``c'' the combined 
component of CDM and baryons, the total matter power 
spectrum, is given by
\be \label{pm}
\langle\delta^m\delta^m\rangle=\left(1-f_{\nu}\right)^2 
\langle \delta^c\delta^c \rangle
+2\left(1-f_{\nu}\right)f_{\nu}
\langle\delta^c\delta^\nu\rangle
+f_{\nu}^2 \langle\delta^\nu\delta^\nu\rangle,
\ee
where $\langle\delta(\vec{k}) \delta(\vec{k}') 
\rangle\equiv (2\pi)^3 \delta_D(\vec{k}+\vec{k}')P(k)$.
If the neutrinos did not induce a gravitational 
backreaction effect on the evolution of the metric and the 
other perturbations, then the maximal effect of the 
neutrino masses would be simply to cut the power spectrum 
by a factor $(1-f_{\nu})^2$ for $k\gg k_{nr}$. However, the 
presence of neutrinos actually modifies the evolution of 
the CDM and baryon density contrasts, such that the linear 
suppression factor is greatly enhanced from 2, roughly by a 
factor of 4, as noted in eq.~\eqref{linsup}. To conclude, 
as also confirmed by numerical simulations the ratio of the 
matter spectra in the massive and massless cases smoothly 
interpolates between 1 for $k<k_{nr}$, and a plateau for 
$k\gg k_{nr}$, as can be seen in the right panel of figure 
\ref{numass}.

\section{Perturbation theory with massive neutrinos}
\label{ptnu}

Our goal in this work is to formulate properly and 
consistently the inclusion of massive neutrinos in the 
evolution of matter perturbations. 

We provide here such formulation using several ingredients. First, we note that 
after the 
non-relativistic transition of neutrinos, the neutrino fraction, $f_\nu$, 
is constant in time, and is very small, in particular in 
light of the current constrained range of total neutrino 
mass in eq.~\eqref{numrange}. Therefore $f_\nu$ is ideal 
for use as the coupling constant of the theory including 
massive neutrinos, and our formulation is made in terms of 
a generic $f_\nu$. This is obviously advantageous in order 
to explore the possible range of this parameter. 
We employ here an exact time evolution, using the proper 
explicit analytic Green's function, common to all $f_{\nu}$ cosmologies, 
rather than resorting to EdS-like approximations. The time evolution can 
be evaluated directly in a simple, sufficiently accurate manner also in 
higher orders than in this work, see e.g.~appendix 
\ref{useapxs}. We also use the fact that the linear neutrino 
component, can be represented to any desirable accuracy as 
a sum of separable functions of $k$, the scale factor 
$a$, and a generic $f_{\nu}$ dependence, with the proper asymptotic 
behavior. Finally, our formulation resides within the general 
perturbative framework of the CDM and baryons as an effective 
fluid 
\cite{Baumann:2010tm,Carrasco:2012cv}, reaching a $1\%$ 
accuracy for small redshifts and larger k modes, that is 
beyond $k\simeq0.1\,h\,Mpc^{-1}$, in the mildly non-linear 
regime. Yet, the ingredients presented in this formulation are generic, and can be equally applied in any perturbation theory.

The massive neutrinos are coupled gravitationally to 
CDM and baryons, and the neutrino density contrast should then 
be specified. After the neutrinos have become fully 
non-relativistic it holds that $\delta^{\nu}\le \delta^c$, with 
an equality at the limit $k\ll k_{nr}$, whereas at $k\gg 
k_{nr}$ we have $\delta^{\nu}\ll \delta^c$. Then the nonlinear 
neutrino component is evaluated here at each order using the 
approximation 
\be \label{clr}
\delta^{\nu}\simeq 
\left(\frac{\delta^{\nu}_{L}}{\delta^{c}_{L}}\right)
\delta^c,
\ee
which indeed maintains the required asymptotic behavior.
This approximation takes into account the backreaction 
effects from non-linearities in the cold dark matter on 
the neutrinos. Yet, we also note that a common simple 
approximation is that the neutrino component remains linear 
throughout time, where indeed all past numerical and 
analytical work support this approximation for the 1\% 
accuracy. In section \ref{exactnlnu} we examine the extent 
to which the inclusion of a nonlinear correction to the neutrino 
component actually affects the leading nonlinear result within the 
1\% precision.

\subsection{The equations of backreaction correction}
\label{eqbr}

In view of the above setup we focus our attention now on the evolution of the effective CDM+baryons component. 
  
Let us begin by writing the density contrast of the 
CDM+baryons component in the following form:
\be\label{defnudec}
\delta^c\equiv\sum_{i=1}^{\infty} 
\delta_i+f_{\nu}\,\sum_{i=1}^{\infty} 
\tilde{\delta}_i\equiv \delta+f_{\nu}\,\tilde{\delta},
\ee
where we denoted $\delta\equiv \sum_{i=1}^{\infty} 
\delta_i$, and 
$\tilde{\delta}\equiv \sum_{i=1}^{\infty} \tilde{\delta}_i$.
This can be regarded as a parametrization 
around $f_{\nu}=0$, where we reduce to the ``bare'' 
perturbation $\delta$ in the massless neutrino case, 
i.e.~$\Lambda$CDM, and $\tilde{\delta}$ represents its 
backreaction correction with the coupling to massive 
neutrinos. Similarly, we rewrite  
the velocity divergence, 
$\theta^c\equiv\vec{\nabla}\cdot\vec{v}^c$, as
\be \label{vfnudec}
\theta^c\equiv\sum_{i=1}^{\infty} 
\theta_i+f_{\nu}\,\sum_{i=1}^{\infty} 
\tilde{\theta}_i,
\ee
and for the neutrino component we just use the usual 
perturbative expansion
$\delta^\nu\equiv \sum_{i=1}^{\infty} 
\delta^\nu_i$.

Let us write the Newtonian equations for subhorizon 
evolution of the CDM+baryons component: 
\begin{align}
\dot{\delta}_c+\frac{1}{a}\partial_j\left[(1+\delta_c)v_c^j
\right]&=0,
\label{ceq}\\
\dot{v}_c^i+Hv_c^i+\frac{1}{a}v_c^j\partial_jv_c^i
+\frac{1}{a}\partial^i\phi&=
-\frac{1}{a}c_s^2\partial^i\delta_c,
\label{nslikeeq}\\
\nabla^2\phi&=\frac{3}{2}\frac{\Omega_m^0 
H_0^2}{a}\left[\left(1-f_{\nu}
\right)\delta_c+f_{\nu}\delta_{\nu}\right],
\label{poisson}
\end{align}
where we have included only the leading EFT counterterm 
\cite{Carrasco:2012cv}, although it will not play a role to the order we 
are considering in this work, and we note that in the Poisson 
equation the CDM+baryons source term is replaced by the 
total matter density contrast as in eq.~\eqref{denmat}, 
including the neutrino component.

Next, we substitute in eqs.~\eqref{ceq}, \eqref{nslikeeq}, 
the Poisson equation, and the decomposition of density 
contrast and velocity divergence from 
eqs.~\eqref{defnudec}, \eqref{vfnudec}, and we write the 
equations independent of $f_\nu$, and linear in $f_\nu$.
Due to the smallness of $f_\nu$ higher orders can be 
neglected. After we Fourier transform the density and 
velocity fields, and assume that the vorticity vanishes, so 
that 
$\vec{v}(\vec{k},t)=-i\frac{\vec{k}}{k^2}\theta(\vec{k},t)$ 
we obtain, as expected, that the evolution equations 
independent of $f_\nu$, are just those of standard 
$\Lambda$CDM for the usual CDM+baryons component with 
massless neutrinos. These equations for $\delta(\vec{k},t)$ 
and $\theta(\vec{k},t)$ 
read:
\begin{align}
a\dot{\delta}+\theta &=-\frac{1}{(2\pi)^3}\int d^3p \left[
\alpha(\vec{p},\vec{k}-\vec{p})\delta(\vec{k}-\vec{p})\theta
\left(\vec{p}
\right)\right],\label{seftc}\\
a\dot{\theta}+aH\theta
+\frac{3}{2}\frac{\Omega_m^0 H_0^2}{a}\delta&=
-\frac{1}{(2\pi)^3}\int d^3p 
\left[\beta(\vec{p},\vec{k}-\vec{p})
\theta(\vec{k}-\vec{p})\theta(\vec{p})\right]+c_s^2k^2\delta
\label{seftns},
\end{align}
where
\begin{align} \label{kernels}
\alpha(\vec{p},\vec{q})=\frac{\left(\vec{p}+\vec{q}\right)
\cdot\vec{p}}{p^2},
\qquad
\beta(\vec{p},\vec{q})=
\frac{1}{2}\frac{\left(\vec{p}+\vec{q}\right)^2
\vec{p}\cdot\vec{q}}{p^2q^2}.
\end{align}

Note that according to our definition in 
eq.~\eqref{defnudec} the linear solution here is expected 
to correspond to that obtained from a linear Boltzmann code
\cite{CAMB,CLASS} for a massless neutrino cosmology.
Further, the merit of this decomposition of the ``bare'' 
perturbations is that the linear growth functions of this 
component are still k independent at the linear level.

Proceeding to the equations linear in $f_{\nu}$ in a 
similar manner, we obtain the following equations for 
$\tilde{\delta}(\vec{k},t)$ and $\tilde{\theta}(\vec{k},t)$:
\begin{align}
a\dot{\tilde{\delta}}+\tilde{\theta}=
-\frac{1}{(2\pi)^3}\int d^3p \left[
\alpha(\vec{p},\vec{k}-\vec{p})\left(\tilde{\delta}(\vec{k}-
\vec{p})
\theta\left(\vec{p}\right)
+\delta(\vec{k}-\vec{p})\tilde{\theta}
\left(\vec{p}\right)\right)\right],\label{fnuc}\\
a\dot{\tilde{\theta}}+aH\tilde{\theta}
+\frac{3}{2}\frac{\Omega_m^0 H_0^2}{a}
\left(\tilde{\delta}+\delta^{\nu}-\delta\right)=
-\frac{2}{(2\pi)^3}\int d^3p 
\left[\beta(\vec{p},\vec{k}-\vec{p})
\theta(\vec{k}-\vec{p})\tilde{\theta}(\vec{p})\right]
+c_s^2k^2\tilde{\delta}.\label{fnunslike}
\end{align}
Notice the unique gravitational source term, which has 
an addition in terms of the difference of density 
components: since after the non-relativistic transition it 
holds that $\delta^\nu\le\delta^c\le\delta$, then we see 
that the free-streaming of massive neutrinos indeed 
gives rise to anti-Gravity for the CDM+baryons component, that is 
the backreaction correction, resulting the suppression of 
growth of CDM+baryons structure formation. 
Also note that due to this additional anti-Gravity source 
term, $\tilde{\delta}$ and $\tilde{\theta}$ have a 
``mixed'' $k$ and time dependence already at the linear 
level, so that at higher orders the nonlinear convolution k integrals 
would actually be more complicated, even though they only 
contain the generic $\Lambda$CDM kernels from 
eq.~\eqref{kernels} in the generic form. This is so since the 
$\delta^\nu$ component, which is a source in the linearized 
equations, has a non-trivial $k$ and time dependence.

We consider the evolution of these last equations as of an 
initial time after the non-relativistic transition, when 
even the lightest neutrinos have become non-relativistic, 
while the non-linearities are still small. Yet, for the 
standard ``bare'' CDM+baryons component in 
eqs.~\eqref{seftc}, by its definition \eqref{seftns} the 
initial time can be extended back to the onset of matter 
domination, or practically to 0.

As we noted, the linear form of these two sets of equations 
is similar up to the additional source term in the 
backreaction equations. Therefore, we resort to the Green's 
function in order to solve them order by order.  
As in standard $\Lambda$CDM we have the following 
homogeneous linear ODE for the linear density contrasts, 
obtained from combining eqs.~\eqref{seftc} and 
\eqref{seftns}, and switching to the scale factor $a$ as 
the independent variable:
\be
-a^4H^2\delta''-(3a^3H^2+a^4HH')\delta'+\frac{3}{2}
\frac{\Omega_m^0 H_0^2}{a}\delta=0.
\ee 
The growing solution of this equation is given by 
\be
\delta_{1+}(a,\vec{k})=C\delta_1(\vec{k})H(a)\int_0^a dx 
\frac{1}{x^3H^3(x)},
\ee
where we define the linear growth function by 
$\delta_{1+}(\vec{k},a)\equiv D_+(a)\delta_1(\vec{k})$,
and $C$ is the normalization constant we fix for the linear 
growth function, e.g.~$C=\frac{5}{2}\Omega_m^0 H_0^2$, or 
that for which $D(a_0=1)=1$, such that
\be \label{gf}
D_+(a)=C H(a)\int_0^a dx \frac{1}{x^3H^3(x)}.
\ee
The decaying solution is given by $D_-(a)=H/H_0$.
Then from eq.~\eqref{seftc}, we have
\be
\theta_1(a,\vec{k})=-a^2H\delta_1'(a,\vec{k})=-a^2H 
D'_+(a)\delta_1(\vec{k}).
\ee

Yet, for the linear correction to the CDM+baryons density contrast 
we have a \textit{nonhomogeneous} linear ODE due to the 
anti-Gravity source coming from the free streaming of 
massive neutrinos. From eqs.~\eqref{fnuc} and 
\eqref{fnunslike} we have:
\be \label{lcdmnh}
-a^4H^2\tilde{\delta}''-(3a^3H^2+a^4HH')\tilde{\delta}'
+\frac{3}{2}\frac{\Omega_m^0 H_0^2}{a}\tilde{\delta}
=\frac{3}{2}\frac{\Omega_m^0 H_0^2}{a}
\left(\delta-\delta^\nu\right).
\ee 
It is essential then to have the Green's function already 
at the linear level. We recall that the Green's 
function should satisfy that 
\be
-a^4H^2(a)\partial^2_aG(a,\bar{a})
-\left[3a^3H^2(a)+a^4H(a)H'(a)\right]\partial_aG(a,\bar{a})
+\frac{3}{2}\frac{\Omega_m^0H_0^2}{a}G(a,\bar{a})
=\delta_D(a-\bar{a}),
\ee 
and that it should hold that $G(a,\bar{a})=0$ for 
$a<\bar{a}$ with the boundary conditions on $a=\bar{a}$ 
being $G(a,\bar{a})|_{a=\bar{a}}=0$, and 
$\partial_aG(a,\bar{a})|_{a=\bar{a}}=
-1/(\bar{a}^4H^2(\bar{a}))$.
We find that the Green's function is given by the following 
analytic closed form:
\be \label{Green}
G(a,\bar{a})=\frac{H(a)}{\bar{a}}\int_a^{\bar{a}} dx 
\frac{1}{x^3H^3(x)} 
\,\theta_H(a-\bar{a}).
\ee
Let us stress that the Hubble parameter, which determines the 
Green's function, corresponds to the total matter here, as 
in the massless neutrino case, and there is no distinction 
here between the CDM+baryons and the neutrino components since 
at this stage we consider the neutrino component to be 
non-relativistic, and thus its background density decays in 
time just as that of CDM+baryons. For this reason this Green's function is common to all $f_\nu$ cosmologies. As the Green's function plays a 
central role in our derivations, we provide an alternative 
efficient accurate evaluation of the Green's function in 
appendix \ref{useapxs} for possible use in higher orders.

\subsection{Leading order backreaction correction}
\label{lobr}

The solution to eq.~\eqref{lcdmnh} of the initial value 
problem is given by
\begin{align} \label{GIVP}
\tilde{\delta}_1(a,\vec{k})=&
\frac{3}{2}\Omega_m^0 H_0^2\int_{a_{in}}^a d\bar{a}
\,\frac{G(a,\bar{a})}{\bar{a}} 
\left[\delta_1(\bar{a},\vec{k})
-\delta^\nu_1(\bar{a},\vec{k})\right]\nn\\
&
+\tilde{\delta}_{in}(\vec{k})
-\tilde{\delta}'_{in}(\vec{k}) a_{in}^4H^2(a_{in}) 
G(a,a_{in}),
\end{align}
where we are considering the initial conditions:
\begin{align}\label{bclin}
\tilde{\delta}_1(a_{in},\vec{k})
=\tilde{\delta}_{in}(\vec{k}),
\qquad
\partial_a\tilde{\delta}_1(a_{in},\vec{k})=
\tilde{\delta}'_{in}(\vec{k}).
\end{align}
and the latter is evaluated numerically from the Boltzmann 
code outputs.

For the velocity divergence we have
\begin{align}
\tilde{\theta}_1(a,\vec{k})= 
-a^2H\left(\frac{3}{2}\Omega_m^0 H_0^2 \int_{a_{in}}^a 
d\bar{a}\,\frac{\partial_a G(a,\bar{a})}{\bar{a}} 
\left[\delta_1(\bar{a})-\delta^\nu_1(\bar{a})\right]
-\tilde{\delta}'_{in}a_{in}^4H^2(a_{in}) 
\partial_a G(a,a_{in})\right).
\end{align}

In the evaluation of these solutions there are two main 
issues to consider. First, is the choice of initial time, 
$a_{in}$. We choose $a_{in}\gtrsim a_{nr}$ according to 
our consideration of the neutrino component becoming fully 
non-relativistic. We recall that this is in fact essential 
within our analysis here in terms of $f_\nu$ as our small 
coupling parameter, which is constant in time only once the 
neutrino component is fully non-relativistic. Yet, $a_{in}$ 
should also be smaller than $a_{NL}$, the time designating the 
onset of significant nonlinear structure formation. Past work
on nonlinear structure formation with massive 
neutrinos contained a wide range of initial redshifts 
$z_{in}=99-9$. Yet, on $z=99$ e.g., 10\% of the neutrinos 
are still relativistic, whereas on $z=9$ nonlinear 
effects are already non negligible. In this work 
we then specify to $z_{in}=24.12$ as the optimal initial time for 
structure formation with massive neutrinos, see also appendix 
\ref{realiz} for further details of our realization. 

\begin{figure}[t]
\begin{centering}
\includegraphics[width=\textwidth]{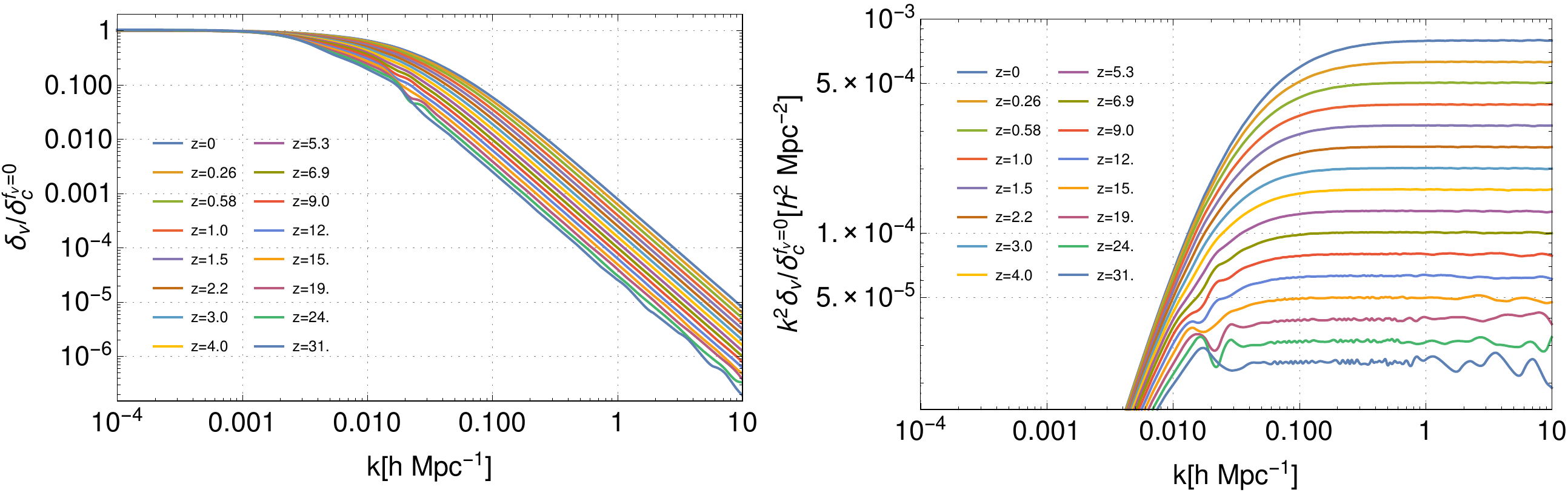}
\caption{The ratio of the linear neutrino component of a 
massive neutrino cosmology with $f_\nu=0.01291$ to the 
CDM+baryons component in the presence of massless 
neutrinos, and its high k asymptotic behavior.}
\label{logfnucombined}
\end{centering}
\end{figure}

Second, we recall that the neutrino component introduces a 
non-trivial ``mixed'' time and $k$ dependence to the source. 
Therefore, we would like to represent it as a sum of separable 
functions of time and $k$, specifically, in the following form:
\begin{align} \label{fitansatz}
\frac{\delta^\nu(a,\vec{k})}{\delta(a,\vec{k})}=\sum_{i=0}^n 
F_{i}(k;f_\nu)\,a^i, \quad n\in N
\end{align}
with as few terms as possible to a desirable accuracy. The 
expansion is done with monomials of the scale factor $a$, being 
the simplest generic time basis. 

\begin{figure}[t]
\begin{centering}
\includegraphics[scale=0.8]{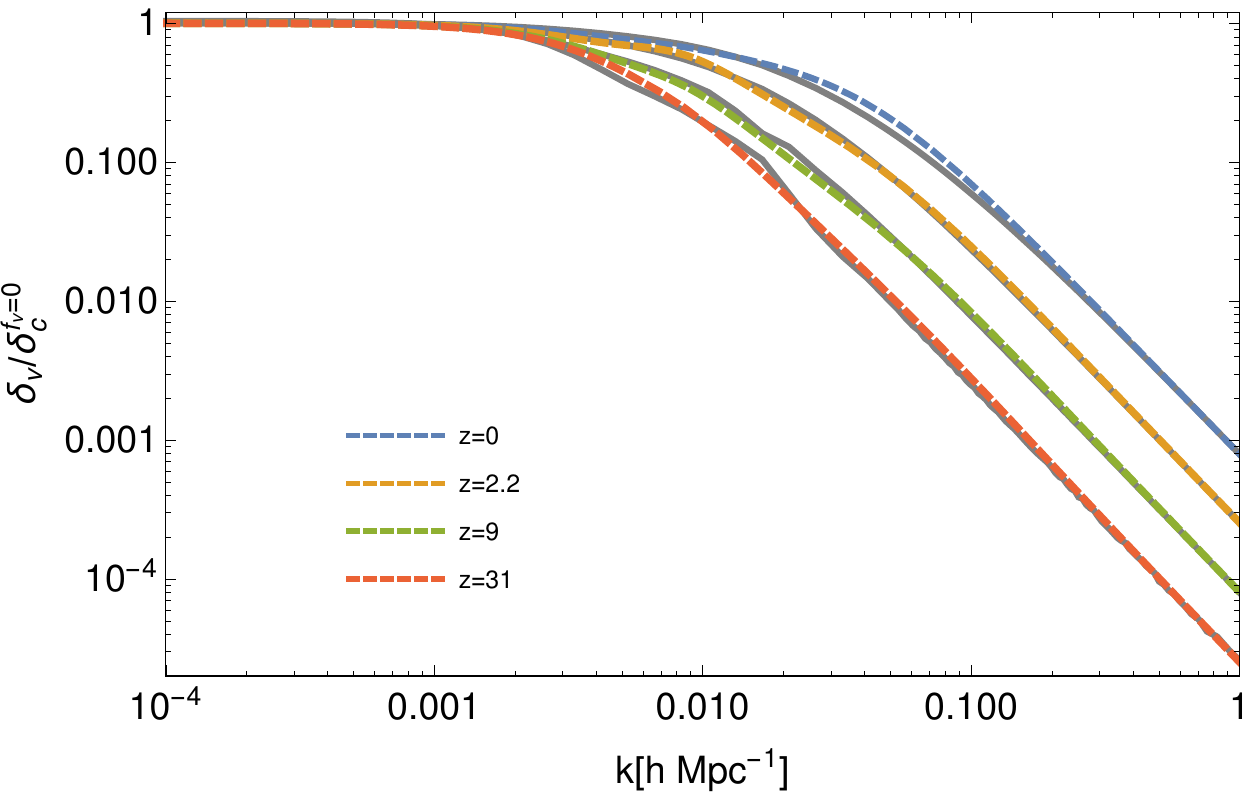}
\caption{The fitting vs.~CLASS data for the ratio 
of the linear neutrino component of a massive neutrino 
Cosmology with $f_\nu=0.01291$ to the CDM+baryons component 
in the presence of massless neutrinos.
The fits at various redshifts are represented by colored 
dashed lines, corresponding to the CLASS data on the gray 
lines.}
\label{fnufit}
\end{centering}
\end{figure}

In figure \ref{logfnucombined} we see this ratio at various 
redshifts, where its asymptotic behavior is found to be 
\be\label{asynu}
\frac{\delta^\nu(a,\vec{k})}{\delta(a,\vec{k})}
\simeq\left\{\begin{array}{ll}
                  1, & \quad k\ll k_{nr},\\
                  Af_\nu^2\,\dfrac{a}{k^2}, & \quad k\gg k_{nr},\\
                \end{array}
              \right.
\ee
with a constant A, and we see that the leading neutrino perturbation spans three orders of 
$f_\nu$ across the k domain.  
Hence, we postulate for $F_i(k;f_{\nu})$ a generic ansatz of 
the form of a fractional function of $k^2$, satisfying the 
leading asymptotic behavior, and make a fitting for the 
free parameters of the ansatz in the proper intermediate k 
domain, $\sim10^{-3}-0.5\, h \,Mpc^{-1}$, using a standard 
least squares fitting procedure. We note that $f_\nu$ is easily 
incorporated as a parameter into the fitting, such that one 
gets a generic $f_\nu$ fitting, and in the following 
we suppress the dependence of the fitting coefficients $F_i$ on 
$f_\nu$. The fitting is shown for various redshifts in 
figure \ref{fnufit}, where in this work three terms were 
included in the expansion in eq.~\eqref{fitansatz}, that is one 
additional term beyond the two necessary to satisfy the 
asymptotics in eq.~\eqref{asynu}, for a permille precision 
in the final results. 

Therefore, with the fitting for the neutrino component in eq.~\eqref{fitansatz} the linear solution in 
eq.~\eqref{GIVP} reads:
\begin{align} \label{denlcor}
\tilde{\delta}_1(a,\vec{k})=\tilde{I}_{1,0}(a)
\delta_1(\vec{k})[1-F_0(\vec{k})]
-\tilde{I}_{1,j}(a)\delta_1(\vec{k})F_j(\vec{k})
+\tilde{\delta}_{in}(\vec{k})-\tilde{\delta}'_{in}(\vec{k}) 
a_{in}^4H^2(a_{in}) G(a,a_{in}),
\end{align}
where we denoted 
\begin{align}
\tilde{I}_{1,j}(a)=
\frac{3}{2}\Omega_m^0 H_0^2\int_{a_{in}}^a d\bar{a}
\,G(a,\bar{a}) D_+(\bar{a})\,\bar{a}^{j-1},
\end{align}
and a summation on the relevant j indices is implied.

\begin{figure}[t]
\begin{centering}
\includegraphics[scale=0.8]{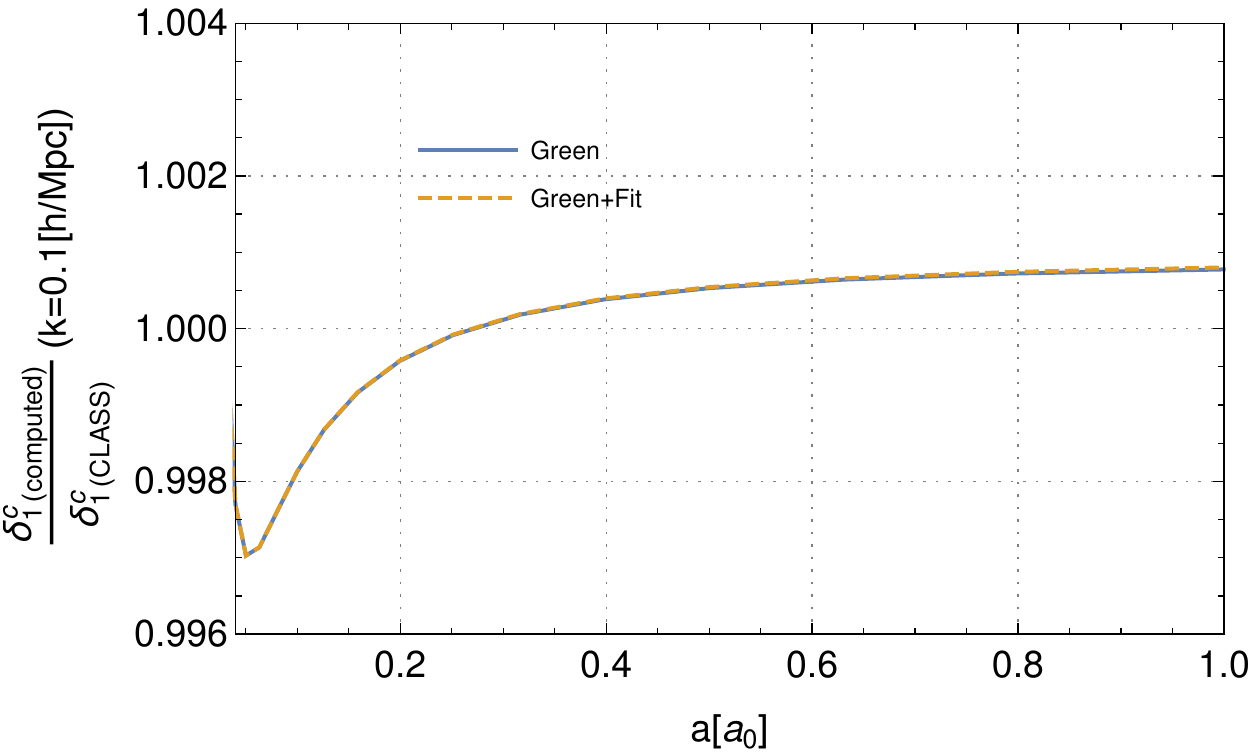}
\caption{The ratio of $\delta_1^c(a)$ computed from Green's 
function, and with fitting of neutrino component, to that 
of CLASS data at k=0.1 h/Mpc in a massive neutrino 
Cosmology of $f_\nu=0.01291$.}
\label{denlin}
\end{centering}
\end{figure}

\begin{figure}[t]
\begin{centering}
\includegraphics[scale=0.8]{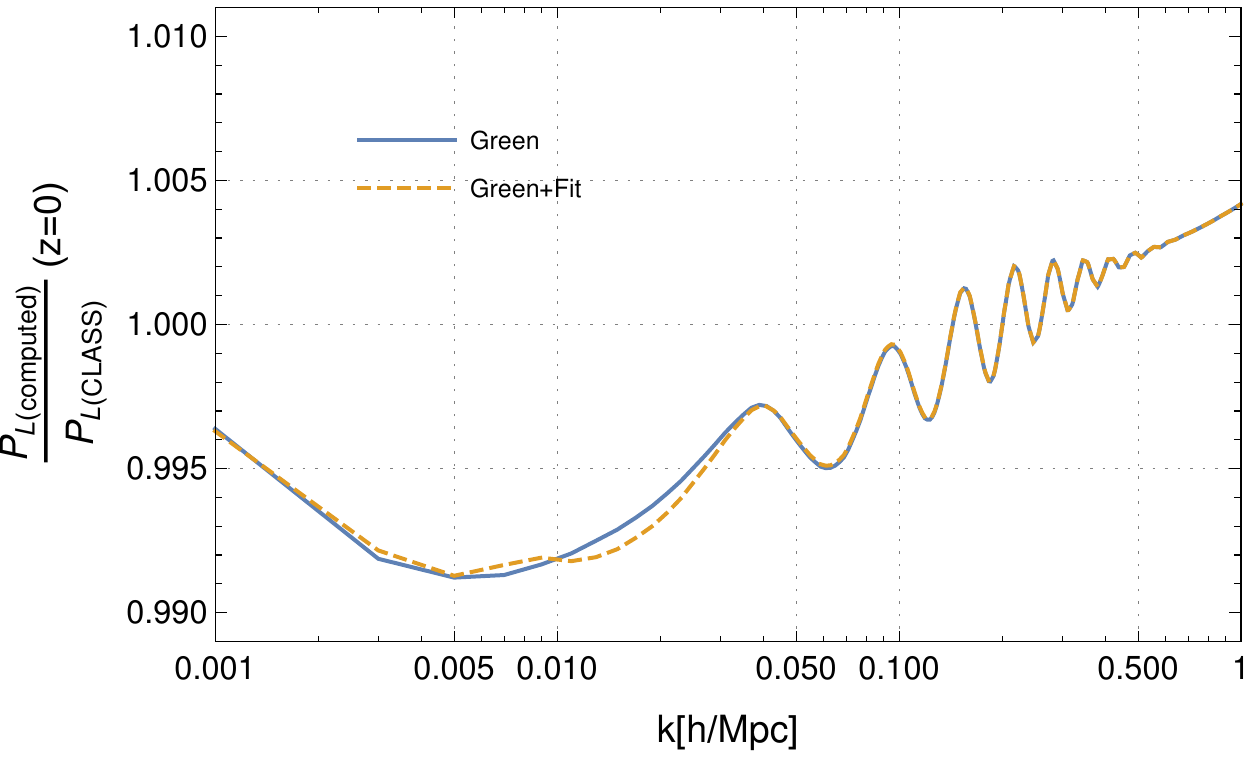}
\caption{The ratio of the linear power spectrum, $P_L$,  
computed from Green's function, and with fitting of 
neutrino component, to that of CLASS data at $z=0$ in a 
massive neutrino Cosmology of $f_\nu=0.01291$.
We note that CLASS data showed a 1\% variance across the 
k domain, and no unique high precision settings to fix 
was possible for this version of the CLASS code (2.4.3).}
\label{complin}
\end{centering}
\end{figure}

In figure \ref{denlin} we show the ratio of the linear 
density contrast computed using eq.~\eqref{GIVP}, 
and using eq.~\eqref{denlcor} with the fitting, to the
that from the CLASS data, and we see an excellent agreement. 
Hence, the fitting allows us to  
include the additional k dependence entering already at 
the linear level in an analytic manner to any desirable 
accuracy, which we can then use in the higher order 
convolution integrals. 

Next, let us consider the linear power spectrum in the 
presence of massive neutrinos. Due to the smallness of the 
neutrino fraction $f_{\nu}$, and since 
$\delta_{\nu}\le\delta$, and $\tilde{\delta}$ is of the 
same order of magnitude as $\delta$, we can write 
\be
\delta_1^m\simeq\delta_1
-f_\nu\left(\delta_1-\tilde{\delta}_1-\delta_1^\nu\right),
\ee
and consider the linear power spectrum to linear order in 
$f_{\nu}$ within the 1\% precision, such that 
\begin{align}
\langle \delta_1^m\,\delta_1^m\rangle\simeq 
(1-2f_\nu)\langle \delta_1^c\,\delta_1^c\rangle
+2f_\nu\langle\delta_1^c\,\delta_1^\nu\rangle
&\simeq(1-2f_\nu)\langle \delta_1\,\delta_1\rangle
+\underbrace{2f_\nu\langle 
\delta_1\,\delta_1^\nu\rangle}_{\text{FS of $\nu$s}}
+\underbrace{2f_\nu\langle 
\delta_1\,\tilde{\delta}_1\rangle}_{\text{BR of CDM}},
\end{align}
where $\langle 
\delta_1(z,\vec{k})\,\delta_1(z,\vec{k}')\rangle\equiv 
(2\pi)^3\delta_D(\vec{k}+\vec{k}')P_L^{f_\nu=0}(z,k)$.
Note the cross term with the neutrino component, which 
vanishes at $k\gg k_{nr}$ due to the free streaming of 
neutrinos, and the cross term of the backreaction 
correction with the ``bare'' CDM+baryons perturbation, 
which represents the leading backreaction effect of 
CDM+baryons, giving rise to the enhanced suppression of 
the linear matter power spectrum in the presence of massive 
neutrinos. The leading backreaction effect of CDM+baryons, 
preceded by a $2f_{\nu}$ factor, explicitly reads
\begin{align}
\frac{\langle \delta_1\,\tilde{\delta}_1\rangle}{\langle 
\delta_1\,\delta_1\rangle}&
=D^{-1}(a)\left[\frac{3}{2}\Omega_m^0 H_0^2 \int_{a_{in}}^a 
d\bar{a}
\, \frac{G(a,\bar{a})}{\bar{a}}D(\bar{a})
\left[1-\sum_{j=0}^n\bar{a}^j F_j(k)\right]\right.\nn\\
&\left.\qquad\qquad\quad+\frac{\tilde{\delta}_i(k)}{\delta_1
(k)}-a_i^4H^2(a_i)G(a,a_i)
\frac{\tilde{\delta}'_i(k)}{\delta_1(k)}\right].
\end{align}

\begin{figure}[t]
\begin{centering}
\includegraphics[scale=0.8]{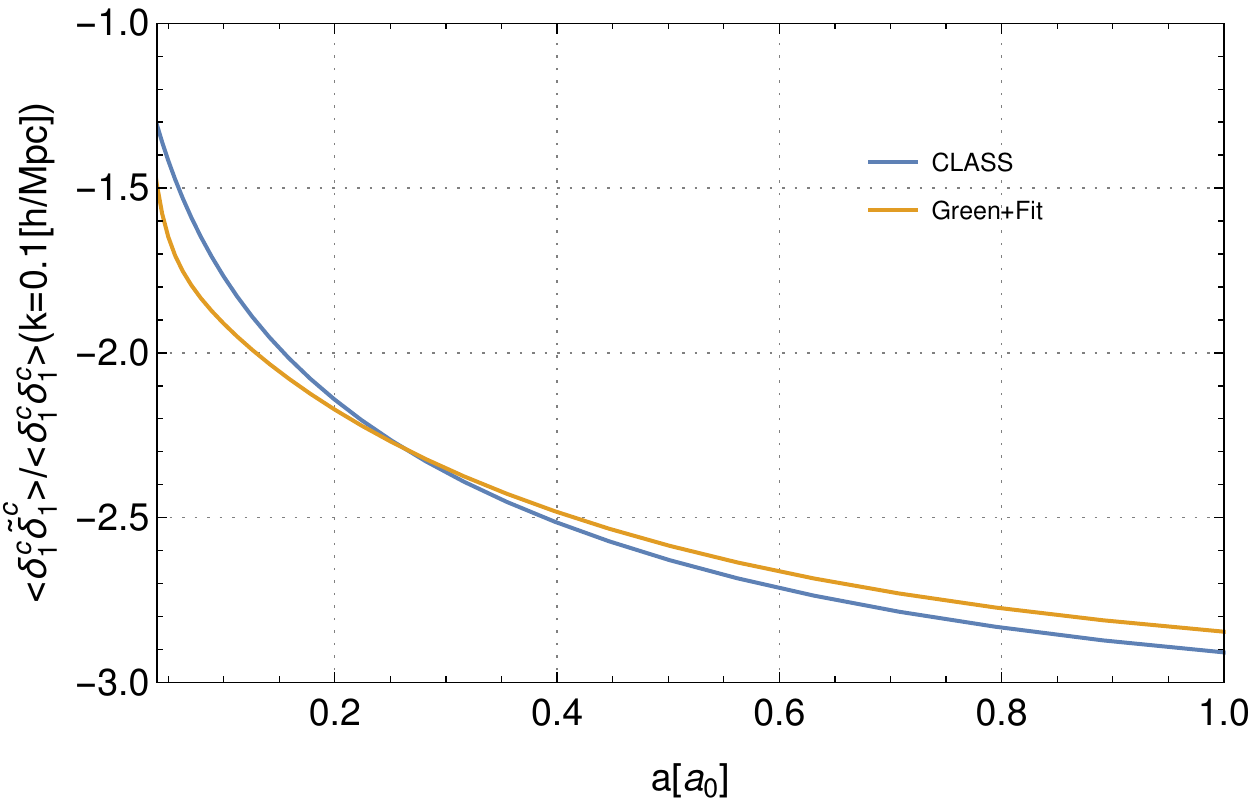}
\caption{The relative effect of the leading CDM+baryons 
backreaction, computed from Green's function with fitting 
of neutrino component, and that of CLASS data 
at $k=0.1 h Mpc^{-1}$ in a massive neutrino cosmology of 
$f_\nu=0.01291$.}
\label{brcdm}
\end{centering}
\end{figure} 

The linear power spectrum at $z=0$ obtained from our 
computation, and with the fitting, vs.~that of the CLASS data is 
shown in figure 
\ref{complin}, whereas the CDM+baryons relative 
backreaction effect at $k=0.1 h Mpc^{-1}$ 
is shown in figure \ref{brcdm}. We can see from both 
figures that our computation is in very good agreement with 
the CLASS data.

\subsection{Leading order matter bispectrum}
\label{lonubis}

For the second order backreaction correction we have to 
compute first the ``bare'' second order 
perturbation of CDM+baryons, $\delta_2$, and from eqs.~\eqref{seftc}, 
\eqref{seftns}, we find 
\begin{align}\label{d2lcdm}
\delta_2(a,\vec{k})=&-\frac{1}{(2\pi)^3}\int d^3p\, 
\delta_1(\vec{p})\delta_1(\vec{k}-\vec{p}) \nn\\
&\qquad\qquad\qquad \times\left[I_1(a)
\left(\alpha(\vec{p},\vec{k}-\vec{p})+\beta(\vec{p},\vec{k}-
\vec{p})\right)
+I_2(a)\alpha(\vec{p},\vec{k}-\vec{p})\right],
\end{align}
where we have 
\begin{align}
I_1(a)=\int_{a_{in}}^a d\bar{a}\, G(a,\bar{a})\bar{a}^4 
H^2(\bar{a})D_{+}^{'2}(\bar{a}),\label{I1}\\
I_2(a)=
\frac{3}{2}H_0^2\Omega_m^0\int_{a_{in}}^a d\bar{a}\, 
G(a,\bar{a})
\frac{D_+^2(\bar{a})}{\bar{a}},\label{I2}
\end{align}
and where we consider the initial conditions:
\begin{align}\label{del2sol}
\delta_2(a_{in}=0,\vec{k})=\delta_2^{in}(\vec{k})=0,
\qquad
\partial_a\delta_2(a_{in}=0,\vec{k})=
\delta_2^{'in}(\vec{k})=0,
\end{align}
and these are fixed at the initial time set at 
EdS, i.e.~$a_{in}=a_{eq}\sim10^{-4}\simeq0$. 

For the second order correction of the CDM+baryons 
perturbation with massive neutrinos, we obtain for 
eqs.~\eqref{fnuc}, \eqref{fnunslike}, the following solution:
\begin{align}\label{d2fnucdm}
\tilde{\delta}_2(a,\vec{k})=
&-\frac{1}{(2\pi)^3}\int d^3p\, 
\delta_1(\vec{p})\delta_1(\vec{k}-\vec{p}) \nn\\
&\qquad\qquad \times\left[
\left(\alpha(\vec{p},\vec{k}-\vec{p})+\beta(\vec{p},\vec{k}-
\vec{p})\right)
\left(\tilde{I}_{2,0}(a)[1-F_0(\vec{k})]
-\tilde{I}_{2,j}(a)F_j(\vec{k})
\right)\right.\nn\\
&\qquad\qquad\qquad+\alpha(\vec{p},\vec{k}-\vec{p})
\left(\tilde{I}_{3,0}(a)
[1-F_0(\vec{k})]-\tilde{I}_{3,j}(a)F_j(\vec{k})\right)\nn\\
&\qquad\qquad\quad+\left(\alpha(\vec{p},\vec{k}-\vec{p})
+\alpha(\vec{k}-\vec{p},\vec{p})
+2\beta(\vec{p},\vec{k}-\vec{p})\right)
\times\nn\\
&\qquad\qquad\qquad
\left(\tilde{I}_{4,0}(a)[1-F_{0}(\vec{p})]
-\tilde{I}_{4,j}(a)
F_{j}(\vec{p})
-\tilde{I}_5(a)
\frac{\tilde{\delta}'_{in}(\vec{p})}{\delta_1(\vec{p})}
\right)\nn\\
&\qquad\qquad\quad+\left(\alpha(\vec{p},\vec{k}-\vec{p})
+\alpha(\vec{k}-\vec{p},\vec{p})\right)\times\nn\\
&\qquad\qquad\qquad\left(I_{6,0}(a)[1-F_{0}(\vec{p})]
-\tilde{I}_{6,j}(a)
F_{j}(\vec{p})+\tilde{I}_{1,0}(a)
\frac{\tilde{\delta}_{in}(\vec{p})}{\delta_1(\vec{p})}
-\tilde{I}_{7}(a)
\frac{\tilde{\delta}'_{in}(\vec{p})}{\delta_1(\vec{p})}
\right)\nn\\
&\qquad\qquad\quad\left.-\alpha(\vec{p},\vec{k}-\vec{p})
\left(\tilde{I}_{1,1}(a)
[1-F_{0}(\vec{p})]-\tilde{I}_{1,j+1}(a)F_{j}(\vec{p})\right)
\right], 
\end{align}
where summation over the proper j indices is implied, and 
we have 
\begin{align}
\tilde{I}_{2,j}(a)=
&\frac{3}{2}\Omega_m^0 H_0^2\int_{a_{in}}^a 
d\bar{a}\,G(a,\bar{a})\,\bar{a}^{j-1} I_1(\bar{a}),\\
\tilde{I}_{3,j}(a)=
&\frac{3}{2}\Omega_m^0 H_0^2\int_{a_{in}}^a 
d\bar{a}\,G(a,\bar{a})\,\bar{a}^{j-1}  I_2(\bar{a}),\\
\tilde{I}_{4,j}(a)=
&\frac{3}{2}\Omega_m^0 H_0^2\int_{a_{in}}^a 
d\bar{a}\,G(a,\bar{a})
\bar{a}^4 H^2(\bar{a}) D'_+(\bar{a})\int_{a_{in}}^{\bar{a}} 
d\hat{a}\,
\partial_{\bar{a}} G(\bar{a},\hat{a}) 
D_+(\hat{a})\,\hat{a}^{j-1},\\
\tilde{I}_5(a)=
& a_{in}^4H^2(a_{in})\int_{a_{in}}^a 
d\bar{a}\,G(a,\bar{a})\bar{a}^4 
H^2(\bar{a})D'_+(\bar{a})\partial_{\bar{a}} 
G(\bar{a},a_{in}),\\
\tilde{I}_{6,j}(a)=
&\frac{3}{2}\Omega_m^0 H_0^2\int_{a_{in}}^a 
d\bar{a}\,G(a,\bar{a})
\frac{D_+(\bar{a})}{\bar{a}}\tilde{I}_{1,j}(\bar{a}),\\
\tilde{I}_7(a)=&\frac{3}{2}\Omega_m^0 H_0^2\, 
a_{in}^4H^2(a_{in})\int_{a_{in}}^a d\bar{a}
\,G(a,\bar{a}) \frac{D_+(\bar{a})}{\bar{a}} 
G(\bar{a},a_{in}),
\end{align} 
and where we consider the initial conditions:
\begin{align}\label{den2cor}
\tilde{\delta}_2(a_{in},\vec{k})
=\tilde{\delta}_2^i(\vec{k})=0,\qquad
\partial_a\tilde{\delta}_2(a_{in},\vec{k})
=\tilde{\delta}_2^{'i}(\vec{k})=0,
\end{align}
and these are fixed at the initial time $a_{in}$, when we 
consider that the neutrinos only start to affect the 
nonlinear structure formation. Note that here we have used 
the approximation in eq.~\eqref{clr} for the second order 
neutrino component.

Let us then consider the LO matter bispectrum in the 
presence of massive neutrinos. 
We recall that the bispectrum is defined as 
\begin{align}
\langle\delta^m(\vec{k}_1,z)\,\delta^m(\vec{k}_2,z)\,
\delta^m(\vec{k}_3,z)\rangle \equiv 
(2\pi)^3\delta_{D}(\vec{k}_1+\vec{k}_2+\vec{k}_3)B(\vec{k}_1,
\vec{k}_2,\vec{k}_3,z).
\end{align}
For the total matter bispectrum we then find, using 
eqs.~\eqref{denmat}, and \eqref{defnudec}, and dropping $f_\nu$ 
beyond linear order within the 1\% precision
\begin{align}\label{bis}
\langle\delta^m(\vec{k}_1)\,\delta^m(\vec{k}_2)\,
\delta^m(\vec{k}_3)\rangle\simeq&
\langle\delta^c(\vec{k}_1)\,\delta^c(\vec{k}_2)\,
\delta^c(\vec{k}_3)\rangle
-f_{\nu}\left[3\langle\delta^c(\vec{k}_1)\,\delta^c(\vec{k}_2)\,
\delta^c(\vec{k}_3)\rangle\right.\nn\\
&\left.-\left(\langle \delta^{\nu}(\vec{k}_1)\,
\delta^c(\vec{k}_2)\,\delta^c(\vec{k}_3)\rangle
+2\,\,  \text{cyc.~perm.}\right) \right]\nn\\
&\simeq\langle\delta(\vec{k}_1)\,\delta(\vec{k}_2)\,
\delta(\vec{k}_3)\rangle
-f_{\nu}\left[3\langle \delta(\vec{k}_1)\,\delta(\vec{k}_2)\,
\delta(\vec{k}_3)\rangle\right.\nn\\
&-\left(\langle \delta^{\nu}(\vec{k}_1)\,
\delta(\vec{k}_2)\,\delta(\vec{k}_3)\rangle
+2\,\,  \text{cyc.~perm.}\right)\nn\\
&\left. -\underbrace{\left(\langle \tilde{\delta}(\vec{k}_1)\,
\delta(\vec{k}_2)\,\delta(\vec{k}_3)\rangle
+2\,\,  \text{cyc.~perm.}\right)}_{\text{backreaction}}\right].
\end{align}
The first non-trivial contribution stems from the first 
nonlinear correction to $\delta_1^m$, i.e.~$\delta_2^m$, which 
yields the tree level bispectrum, with the term that reads:
\begin{align}\label{bisapp}
\langle &\delta_1^m(\vec{k}_1)\,\delta_1^m(\vec{k}_2)\,
\delta_2^m(\vec{k}_3)\rangle\simeq 
\langle\delta_1(\vec{k}_1)\,\delta_1(\vec{k}_2)\,
\delta_2(\vec{k}_3)\rangle\left(1-3f_\nu\right)\nn\\
&+f_\nu\left(\langle \delta_1^{\nu}(\vec{k}_1)\,
\delta_1(\vec{k}_2)\,\delta_2(\vec{k}_3)\rangle
+\langle \delta_1(\vec{k}_1)\,
\delta_1^{\nu}(\vec{k}_2)\,\delta_2(\vec{k}_3)\rangle
+\frac{\delta_1^{\nu}(k_3)}{\delta_1(k_3)}
\langle \delta_1(\vec{k}_1)\,\delta_1(\vec{k}_2)\,
\delta_2(\vec{k}_3)\rangle \right)\nn\\
&+f_\nu\left(\langle \tilde{\delta}_1(\vec{k}_1)\,
\delta_1(\vec{k}_2)\,\delta_2(\vec{k}_3)\rangle
+\langle \delta_1(\vec{k}_1)\,
\tilde{\delta}_1(\vec{k}_2)\,\delta_2(\vec{k}_3)\rangle
+\langle \delta_1(\vec{k}_1)\,
\delta_1(\vec{k}_2)\,\tilde{\delta}_2(\vec{k}_3)\rangle \right),
\end{align}
and we have made use of the approximation in eq.~\eqref{clr} 
to evaluate the term with the second order neutrino 
perturbation, such that:
\begin{align}\label{nlnuexplicit}
\langle\delta_1^{c}(\vec{k}_1)\,\delta_1^c(\vec{k}_2)\,
\delta_2^{\nu}(\vec{k}_3)\rangle=
\langle\delta_1^{c}(\vec{k}_1)\,\delta_1^c(\vec{k}_2)\,
\delta_2^{c}(\vec{k}_3)\rangle
\frac{\delta_1^{\nu}(k_3)}{\delta_1^{c}(k_3)},
\end{align}
and dropped higher orders in $f_\nu$. Notice that the first two 
terms with the linear neutrino component in eq.~\eqref{bis} can 
then be brought to a similar form, of the massless neutrino 
bispectrum times the ratio of neutrino to CDM components.

\begin{figure}[t]
\begin{centering}
\includegraphics[width=\textwidth]{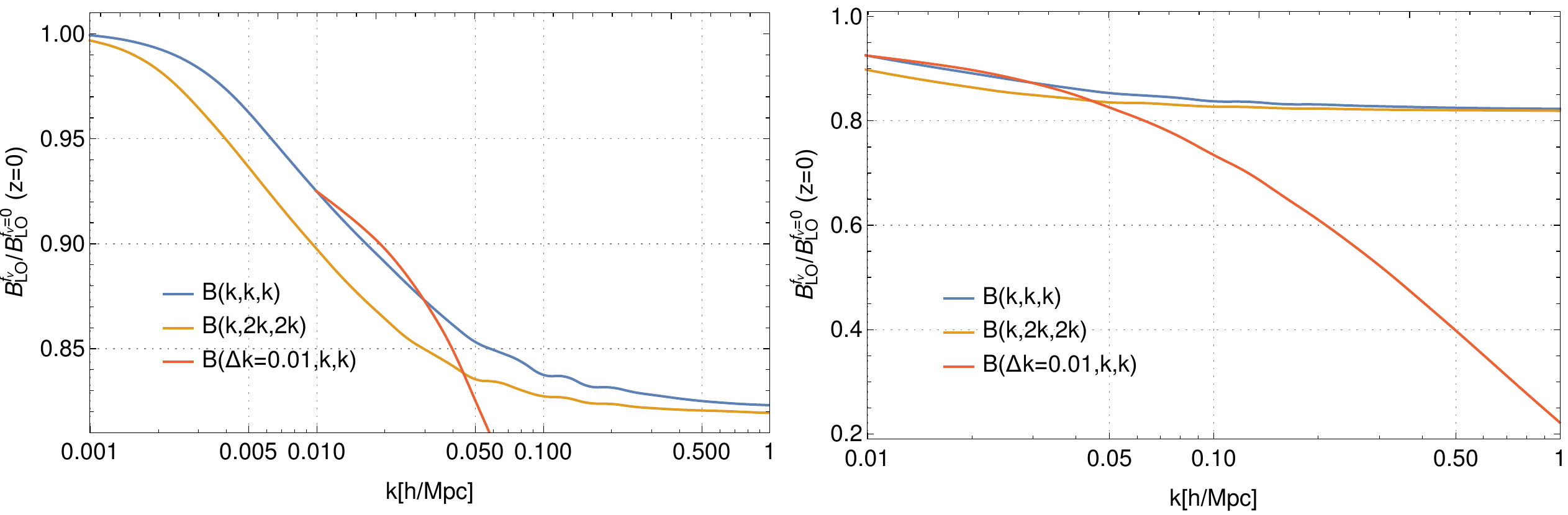}
\caption{The ratio of LO matter bispectra  with 
$f_\nu=0.01291$ to $f_\nu=0$ at $z=0$ 
of three special configurations. The squeezed limit 
configuration shows a significant suppression at 
high k modes, suggesting the squeezed limit as a sensitive 
probe of neutrino mass.} 
\label{bis2d}
\end{centering}
\end{figure}

We recall that the LO bispectrum in EdS reads:
\begin{align}
B_{LO}^{EdS}(\vec{k}_1,\vec{k}_2,\vec{k}_3,z)=
&\left(\frac{5}{7}\left[\alpha(\vec{k_1},\vec{k_2})
+\alpha(\vec{k_2},\vec{k_1})\right]
+\frac{4}{7}\beta(\vec{k_1},\vec{k_2})\right)
P_L(k_1,z)P_L(k_2,z)\nn\\
&+2\,\,\text{cyc.~perm.},
\end{align}
and for $\Lambda$CDM, or the massless neutrino case, we get from 
eq.~\eqref{d2lcdm} that the LO bispectrum reads:
\begin{align}\label{bislcdm}
B_{LO}^{f_{\nu}=0}(\vec{k}_1,\vec{k}_2,\vec{k}_3,z)=&
-\left(\frac{I_1(a)}{D^2_+(a)}\left[\alpha(\vec{k_1},\vec{k_2})
+\alpha(\vec{k_2},\vec{k_1})+2\beta(\vec{k_1},\vec{k_2})\right]
\right.\nn\\
&\left.\quad+\frac{I_2(a)}{D^2_+(a)}
\left[\alpha(\vec{k_1},\vec{k_2})+\alpha(\vec{k_2},\vec{k_1})
\right]\right)P_L(k_1,z)P_L(k_2,z)\nn\\
&+2\,\,\text{cyc.~perm.}
\end{align}
In order to explore the shape dependence of the bispectrum a 
reduced bispectrum can be defined as
\be
Q=\frac{B(\vec{k_1},\vec{k_2},\vec{k_3})}
{P_L(k_1)P_L(k_2)+P_L(k_2)P_L(k_3)+P_L(k_3)P_L(k_1)}.
\ee
We then consider $Q$ for a fixed $k_1$ as 
a function of $x_2=k_2/k_1$ and $x_3=k_3/k_1$, which are 
constrained to $x_2\ge x_3\ge 1-x_2$.
We show the ratio of reduced bispectra of EdS and $\Lambda$CDM 
in figure \ref{lcdmbis} in appendix \ref{useapxs}. 

\begin{figure}[t]
\begin{centering}
\includegraphics[scale=0.6]{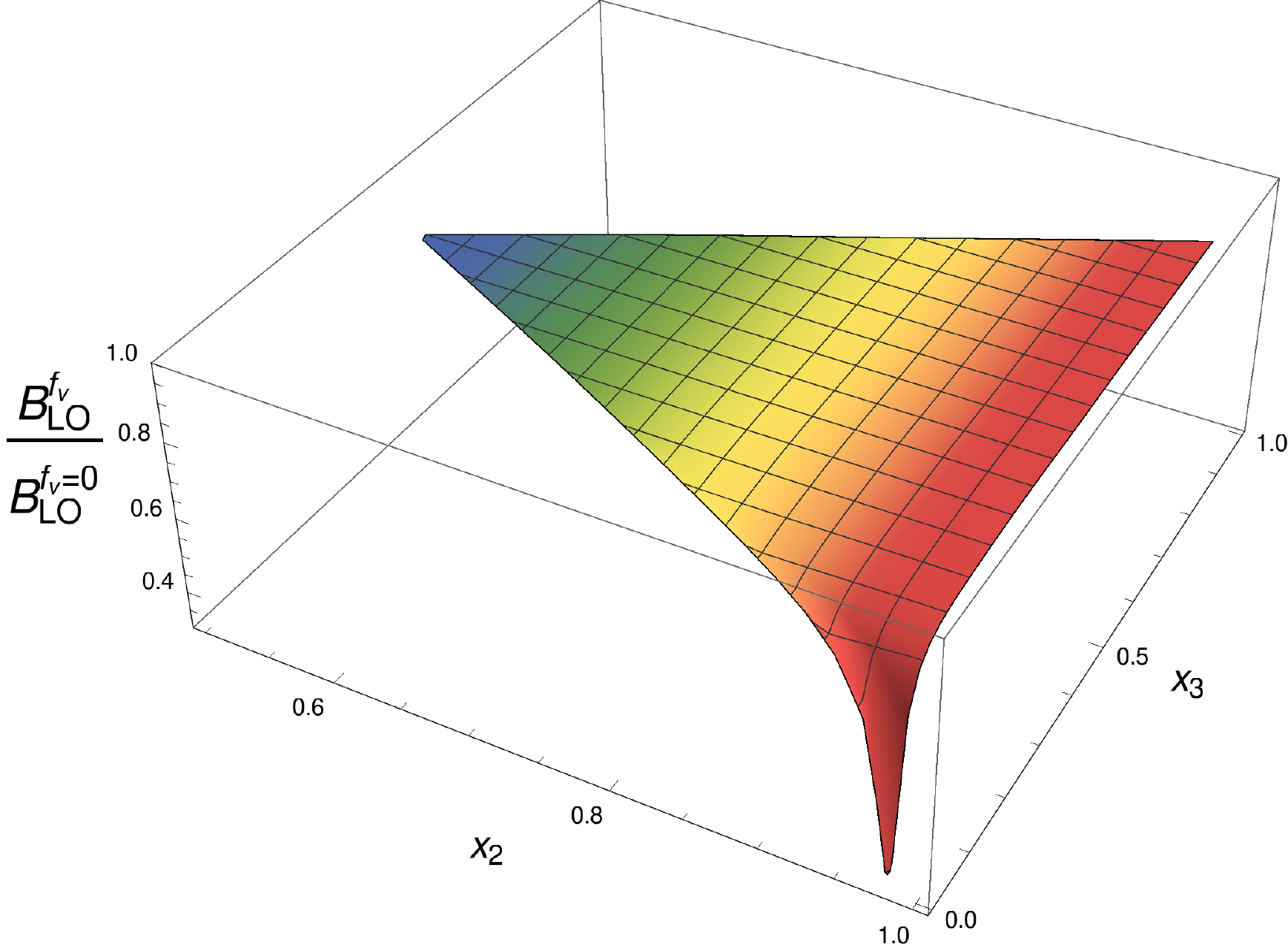}
\caption{The ratio of LO matter bispectra with 
$f_\nu=0.01291$ to $f_\nu=0$ at $z=0$. 
We consider the shape dependence for a fixed $k_1=0.2\,h\,Mpc^{-1}$ as 
a function of $x_2=k_2/k_1$ and $x_3=k_3/k_1$, which are 
constrained to $x_2\ge x_3\ge 1-x_2$.
The shape dependence shows a suppression of similar values to the 
equilateral configuration suppression of $\sim-13.5f_{\nu}$ , whereas 
a steep enhanced suppression appears around the squeezed limit at high k 
modes.}
\label{bis3d}
\end{centering}
\end{figure}

We examine here the particular cases of equilateral configuration, 
i.e.~$k_1=k_2=k_3$, the triangle configuration given by 
$k_2/k_1=k_3/k_1=2$, and the squeezed limit, i.e.~$k_1\to0$, 
$k_2/k_3=1$. These bispectra are seen in figure \ref{bis2d}.
The tree level bispectrum should be similar to SPT extensions 
as in \cite{Fuhrer:2014zka}, since there are still no counterterms 
contributing at this perturbative order.   
For the equilateral configuration we find that the suppression 
of the LO bispectrum in the presence of massive neutrinos is 
given by
\be
\frac{\Delta 
B_{LO}^{f_\nu}(k,k,k)}{B_{LO}^{f_\nu=0}(k,k,k)}\simeq-13.5\,f_\nu,
\ee
in analogy to eq.~\eqref{linsup} for the suppression of the 
linear power spectrum. Hence, the suppression effect that we 
find here is slightly larger than that estimated in  
\cite{Fuhrer:2014zka}, which started the evolution at a notably 
late redshift, where nonlinear effects cannot be neglected.
The reduced bispectrum for a fixed $k_1=0.2 \,h\,Mpc^{-1}$ is seen in 
figure \ref{bis3d}, where we see  that the shape dependence shows a 
suppression of similar values to the equilateral configuration 
suppression of $\sim-13.5f_{\nu}$ , whereas a steep enhanced suppression 
appears around the squeezed limit at high k modes.

\subsection{Relevance of exact evaluation of nonlinear neutrino perturbation} 
\label{exactnlnu}

\begin{figure}[t]
\begin{centering}
\includegraphics[scale=0.6]{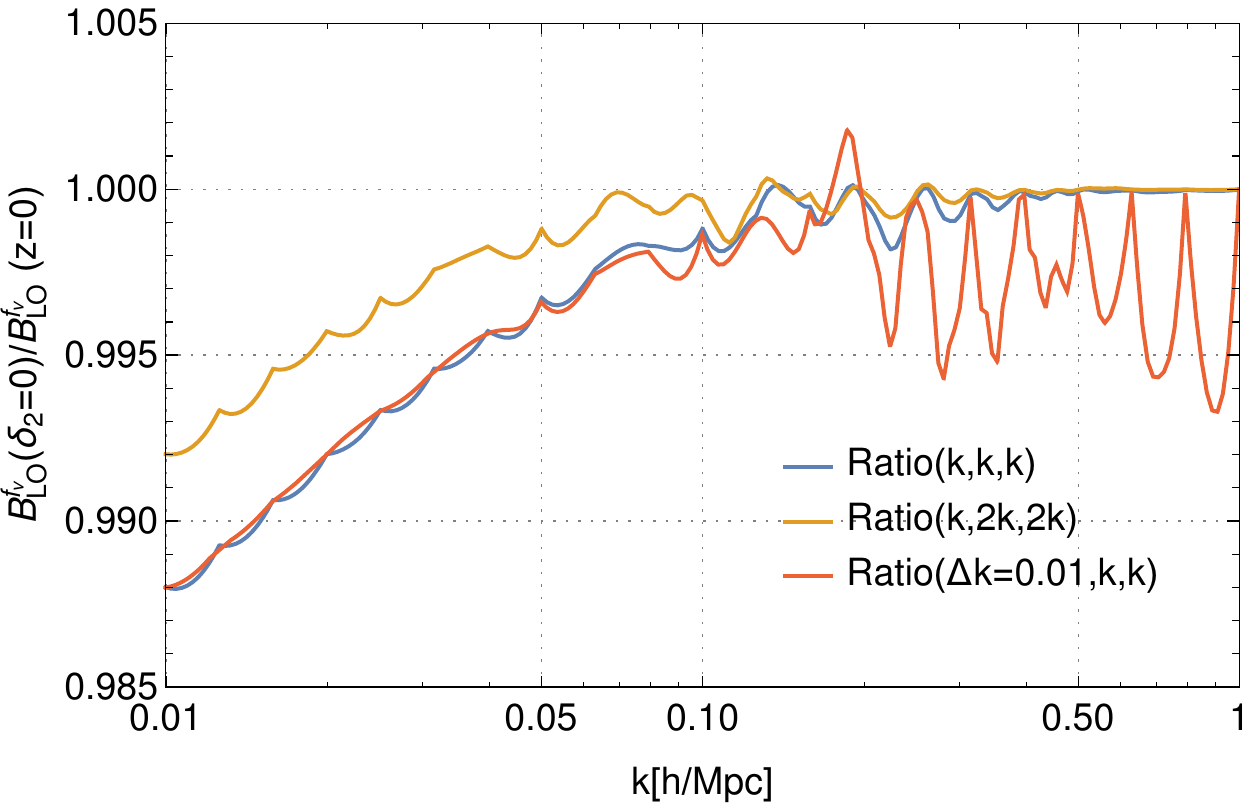}
\caption{The ratio of LO matter bispectra with 
$f_\nu=0.01291$, and $\delta_2^{\nu}=0$ to the bispectra computed in 
this work, using eq.~\eqref{clr}, at $z=0$. 
We can see that even a wider range of possible values for the exact 
nonlinear neutrino perturbation leads to a range of results, which is  
within the 1\% precision for practically all observable scales.}
\label{bis2dcheck}
\end{centering}
\end{figure}

Let us reconsider now the relevance of the exact evaluation of the 
nonlinear neutrino perturbation for the desirable 1\% precision in 
nonlinear perturbation theory with massive neutrinos.
We recall that in this work we have used for the nonlinear neutrino 
perturbation the evaluation given by eq.~\eqref{clr}.
This is in fact an upper bound for the exact NL value, since the NL 
clustering of neutrinos is expected to be even more suppressed, compared 
to that of CDM+baryons, than their relative linear clustering. 
On the other hand, an extremely crude lower bound would be just to take 
the nonlinear neutrino perturbation to vanish. Hence, the actual exact 
nonlinear neutrino perturbation is found in the following possible crude range
\be\label{nunlrange}
0<\delta_{NL}^{\nu}\lesssim
\left(\frac{\delta^{\nu}_{L}}{\delta^{c}_{L}}\right)
\delta^c_{NL}, 
\ee
and therefore an overly critical way to examine the importance of the 
exact evaluation of the nonlinear neutrino perturbation, would be to 
compare the results we would have obtained by taking the extreme lower 
bound, that is by assuming the nonlinear neutrino perturbation vanishes, 
and compare them with our results. We note that in this work the NL 
neutrino perturbation enters the final result from two origins: 
First in the anti-Gravity source for the NL CDM+baryons backreaction 
correction, i.e.~in eq.~\eqref{d2fnucdm}. Second, explicitly in the 
bispectrum cross correlation of neutrino with CDM+baryons in eq.~ 
\eqref{nlnuexplicit}. The results of this check can be seen in figure 
\ref{bis2dcheck}. We can see that even the extreme range of possible 
values for the exact nonlinear neutrino perturbation leads to a range of 
results, which is within the 1\% precision for practically all 
observable scales.

For completeness we note previous approaches to estimate the effect of 
the nonlinear neutrino perturbation on nonlinear results. In 
\cite{Saito:2008bp,Saito:2009ah} the neutrino perturbations were 
numerically approximated by solving the \textit{modified} linearized 
Boltzmann equation, using the Boltzmann code CAMB \cite{CAMB}, via
plugging into the Poisson equation approximations of the NL density 
contrasts of CDM+baryons. The effect of this estimate was found there 
well below the 1\% precision, and hence the linear neutrino assumption 
was adopted. In \cite{Blas:2014hya} the nonlinear neutrino perturbation 
was evaluated by treating the neutrino component as a fluid. Yet, as we 
already stressed above, this provides a poor description for the 
free streaming neutrino component on all relevant observable scales, 
which are larger than the characteristic free streaming scale. Finally, 
other approaches to evaluate the nonlinear neutrino component were put 
forward in \cite{Dupuy:2013jaa}, which requires the explicit evaluation 
of the perturbed neutrino momentum distribution in real time, and in 
\cite{Fuhrer:2014zka}, which is also computationally intensive, making 
both approaches impractical. As our results in figure \ref{bis2dcheck} 
illustrate, along with similar conclusions of all previous analytical
and numerical studies of NL observables with massive neutrinos, the 
exact evaluation of the NL neutrino perturbation, has negligible
impact on the final results, at least up to the two-loop level, and 
therefore does not constitute a crucial aspect in evaluating 
successfully to 1\% precision NL LSS with massive neutrinos. It is 
rather the exact evaluation of the CDM+baryons perturbations, and their 
backreaction corrections, as is realized in this work.

\section{Conclusions}
\label{theendmyfriend}

In this paper we introduced a consistent formulation for a 
perturbation theory that incorporates massive neutrinos in the 
evolution of matter perturbations, such that all linear k dependence 
in the growth functions of CDM+baryons perturbations, as well as all 
consequent additional mode coupling at higher orders are taken into 
account to any desirable accuracy. This is achieved using the fact that 
the linear neutrino perturbation, can be represented to any 
desirable accuracy as a sum of separable functions of $k$, 
the scale factor $a$, and a generic $f_{\nu}$ dependence, with the 
proper asymptotic behavior. Our formulation is based on that after the 
non-relativistic transition of the neutrinos, the neutrino 
fraction, $f_\nu$, is constant in time, and much smaller than 
unity, in particular in light of the current constraints on 
the total neutrino mass. Therefore, $f_\nu$ is regarded as the coupling 
constant of the theory including massive neutrinos, and our formulation 
is made in terms of a generic $f_\nu$, which is clearly advantageous 
for exploring the possible range of this particle physics 
parameter. Then the ``bare'' perturbations are those in the 
massless neutrino case when the neutrino fraction vanishes, and 
we consider the backreaction corrections due to the gravitational 
coupling of massive neutrinos. This also allows to consider the 
``bare'' perturbations nonlinearities from an earlier redshift than 
their backreaction corrections, and hence capture dominant nonlinear 
effects, which occur before neutrinos become fully 
non-relativistic. 

We have derived the general equations for the 
``bare'' perturbations and their backreaction corrections, and carried 
out the exact time evolution, using the proper explicit analytic 
Green's function, common to all $f_\nu$ cosmologies. We explicitly 
derived the leading backreaction effect, and found precise agreement with the 
linear effect, and the linear total matter power spectrum. 
Furthermore, we computed the second order backreaction correction, 
and derived the leading order matter bispectrum in the presence 
of massive neutrinos, suggesting that the squeezed limit of the LO matter bispectrum as a sensitive probe of neutrino mass.
Finally, we have also demonstrated the irrelevance of the exact 
evaluation of the nonlinear neutrino perturbation for the 1\% precision 
of our NL result in agreement with the conclusions of all previous 
studies of NL LSS with massive neutrinos.

Our perturbation theory resides within the general view of the CDM and 
baryons as an effective 
fluid, and is therefore well-defined, extending to a larger k 
reach in the mildly non-linear regime. Yet the generic formulation 
in this work allows for the consistent inclusion of massive 
neutrinos within any perturbative theory. We have employed an 
exact time evolution, using the analytic Green's function 
similar to that of the baseline $\Lambda$CDM cosmology. Thus we 
refrained from resorting to EdS-like approximations, 
while still maintaining computational efficiency for 
practical use. The relevance of the exact time evolution for 
sub-percent precision calculations has also been recently investigated 
in \cite{Fasiello:2016qpn}, where it has been pointed out that these 
EdS-like approximations are less suited for higher order velocity 
statistics, relevant to all observables affected by redshift space 
distortion effects.

In this respect the only practical challenge in our 
formulation, that has to be tackled at higher orders, is the 
multiple time integrations, increasing with each order of the 
perturbative theory. For this purpose the generic time 
integrals of the Green's function with powers of the scale 
factor should be considered for a possible analytical and 
useful accurate numerical simplification, such as those noted 
in appendix \ref{useapxs}. Further, the time integrations may 
be numerically implemented in non-trivial ways for more efficient  
evaluation. Finally, we note that external robust integration 
routines such as the CUBA library \cite{CUBA} may also be very 
useful to increase the efficient evaluation performance.
We leave it for future work to extend the implementation of our 
formulation to higher orders of n point functions, and to be 
confronted with N-body simulations.

\acknowledgments

ML thanks Leonardo Senatore for collaboration, and kind hospitality at 
the Stanford Institute of Theoretical Physics, in early stages of this 
work. 
ML also thanks Simon Foreman for useful discussions.
ML is supported by the European Research Council under the European 
Community's Seventh Framework Programme (FP7/2007-2013 Grant Agreement 
No.~307934, NIRG project). 
This work has been done within the LABEX ILP (reference ANR-10-LABX-63) 
part of the Idex SUPER, and received  financial French state aid managed 
by the Agence Nationale de la Recherche, as part of the programme 
Investissements d'Avenir under the reference ANR-11-IDEX-0004-02.
ZV is supported in part by the U.S.~Department of Energy 
contract to SLAC No.~DE-AC02-76SF00515.

\appendix

\section{Useful numerical approximations} 
\label{useapxs}

\begin{figure}[t]
\begin{centering}
\includegraphics[width=\textwidth]{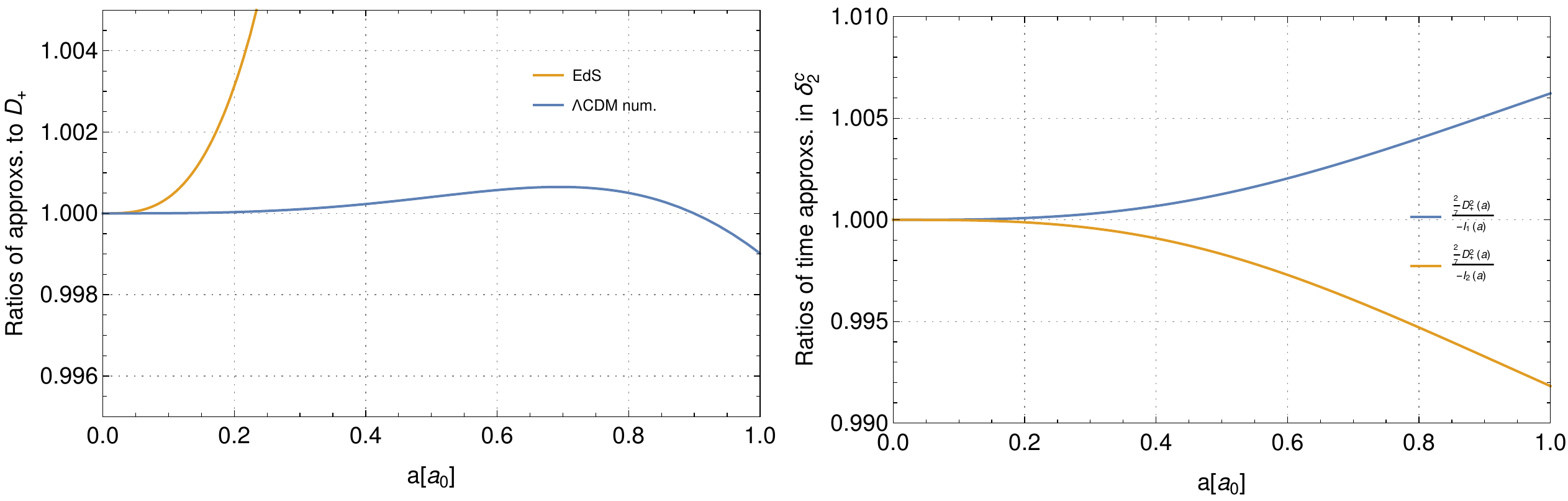}
\caption{\textit{Left:} The ratios of the EdS approximation 
and of the approximation in \eqref{gfapp} to the linear 
growth function $D_+$. \textit{Right:} The ratios of the 
EdS-like approximation in eq.~\eqref{edslikeapx} to the 
time integrals, appearing in the second order density 
perturbation in $\Lambda$CDM.}
\label{apxscombined}
\end{centering}
\end{figure}

For $\Lambda$CDM the growth function given in \eqref{gf} 
can be approximated to 1 permille precision, using the 
function given by 
\cite{Bernardeau:2001qr}
\be \label{gfapp}
D_+(a)\simeq N d(a)= N\,
\frac{a\Omega_m}{\Omega_m^{4/7}
-\Omega_{\Lambda}+\left(1+\frac{\Omega_m}{2}
\right)\left(1+\frac{\Omega_{\Lambda}}{70}\right)},
\ee
where N is the proper normalization constant. This 
approximation is shown in figure \ref{apxscombined}.

The Green's function in eq.~\eqref{Green} can be written 
directly in terms of the growth factor and the Hubble 
parameter as follows:
\begin{align}
G(a,\bar{a})=C^{-1}\frac{D_+(\bar{a})}{\bar{a}}\left(\frac{D_-(a)}{D_-(\bar{a})}
-\frac{D_+(a)}{D_+(\bar{a})}\right)\,\theta_H(a-\bar{a}),
\end{align}
where $C$ is the normalization constant of the growth factor from eq.~\eqref{gf}, and hence eq.~\eqref{gfapp} can also be used in the Green's 
function for an efficient numerical evaluation, rather than 
numerically evaluating the time integral in 
eq.~\eqref{Green}. 

\begin{figure}[t]
\begin{centering}
\includegraphics[scale=0.5]{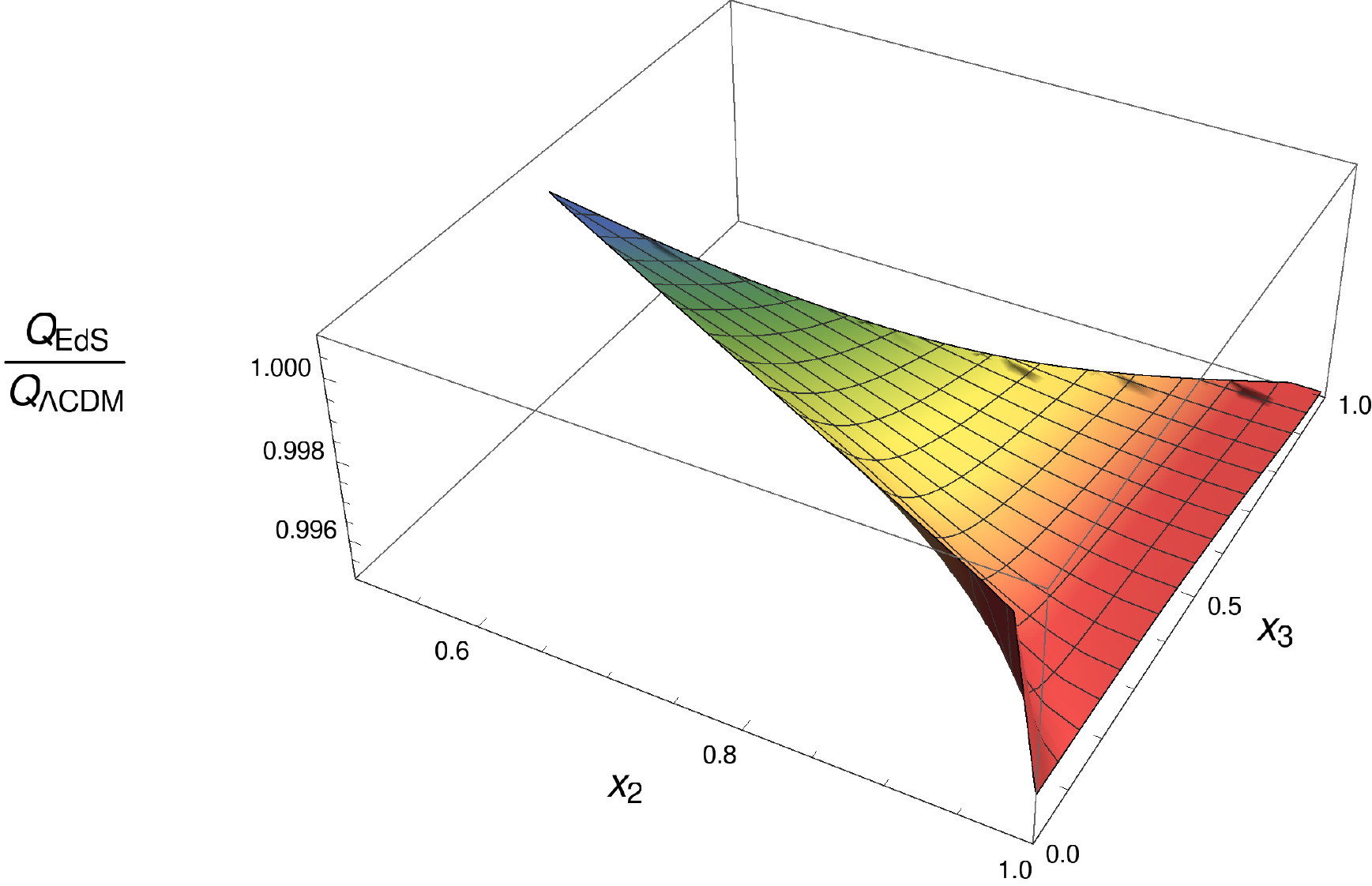}
\caption{The ratio of the reduced bispectra of the EdS-like 
approximation to $\Lambda$CDM, demonstrating the 1\% 
precision attained, using the EdS-like approximation in 
eq.~\eqref{edslikeapx}, rather than the exact Green's time 
integral. The reduced bispectra are presented for a fixed 
$k_1=0.2\, h \,Mpc^{-1}$ as a function of $x_2=k_2/k_1$ and 
$x_3=k_3/k_1$, which are constrained to $x_2\ge x_3\ge 1-x_2$. }
\label{lcdmbis}
\end{centering}
\end{figure}

The time integrals from eqs.~\eqref{I1}, 
\eqref{I2}, appearing in the second order standard 
$\Lambda$CDM density perturbation in eq.~\eqref{d2lcdm}, 
and consequently in the $\Lambda$CDM bispectrum in 
eq.~\eqref{bislcdm} can be approximated within the 
1\% precision by their EdS-like value, that is
\be\label{edslikeapx}
I_1(a)\simeq I_2(a)\simeq-\frac{2}{7}D_+^2(a),
\ee
as can be seen in figure \ref{apxscombined}. The effect of 
using this EdS-like approximation in the LO bispectrum of 
$\Lambda$CDM is shown in figure \ref{lcdmbis}.

We should stress though that none of these approximations 
was employed in this work, where the exact computations 
were still very manageable. 

\section{Realization}
\label{realiz}

In this work we have used the recent Boltzmann C code CLASS 
(version 2.4.3) \cite{Blas:2011rf, CLASS} to set up the 
initial conditions, e.g.~to compute the input transfer 
functions for the CDM, baryons, and neutrino components in 
the massless and massive cases. CLASS seems to be better 
suited for the inclusion of massive neutrinos, which is 
implemented differently than in the commonly used CAMB code 
\cite{Lesgourgues:2011rh}.

The main input parameters which were used for all neutrino
cosmologies, following Planck 2013 \cite{Ade:2013zuv}, are 
noted in table \ref{cospars}. Additional input parameters 
which were used for all cosmologies are: Photon density 
$T_{\text{cmb}} = 2.726$, primordial Helium fraction $YHe = 
0.25$, pivot scale in $Mpc^{-1}$ $k_{pivot} = 0.05$, and 
tilt running $\alpha_s = 0$. The input parameters, which 
varied among the neutrino cosmologies, in particular 
between the massless and massive cases, appear in table 
\ref{nupars}. For the massive case we considered a single 
distinct species with 3 degenerate massive neutrinos.

\begin{table}[t]
\begin{centering}
\begin{tabular}{|c|c|c|c|c|} 
\hline
h & $\Omega_m$ & $\Omega_b$ & $A_s$ & $n_s$\\
\hline\hline
0.67 & 0.3167 & 0.05 & 2.1265$\times10^{-9}$ & 0.96\\
\hline
\end{tabular}
\caption{The cosmological parameters used in our massless 
and massive neutrino realizations.}
\label{cospars}
\end{centering}
\end{table}

\begin{table}[t]
\begin{centering}
\begin{tabular}{|l|c|c|c|c|} 
\hline
$M_\nu$[eV] & $f_\nu$ & $\Omega_{\text{cdm}}$& $N_{\textbf{eff}}$& $N_{ncdm}$ \\
\hline\hline
0.0 & 0 & 0.2667 & 3.04 & ---\\
\hline
0.06 & 4.531$\times10^{-3}$ & 0.2653 & 0.0 & 1\\
\hline
0.102 & 7.703$\times10^{-3}$ & 0.2643 & 0.0 & 1\\
\hline
0.141 & 1.065$\times10^{-2}$ & 0.2633 & 0.0 & 1\\
\hline
0.171 & 1.291$\times10^{-2}$ & 0.2626 & 0.0 & 1\\
\hline
\end{tabular}
\caption{The various neutrino cosmologies parameters used 
in our realizations.}
\label{nupars}
\end{centering}
\end{table}

We output the matter transfer functions. In particular, we 
get the density contrast for each \textit{non-degenerate} 
non-CDM species, hence in our case we get one value, and we 
consider the neutrino density fixed and distributed equally 
amongst three massive neutrino degenerate species.  

The k domain of output in units of $h\,Mpc^{-1}$ is $\sim 
10^{-5}-10^3$, and we took 15 uniform log intervals of 
redshifts, such that $Log[1+z]=0,..,1.4$, corresponding to 
the redshifts $z=$0, 0.26, 0.58, 1.00, 1.51, 2.16, 2.98, 
4.01, 5.31, 6.94, 9.00, 11.59, 14.85, 18.95, 24.12.

Finally, we note that no unique high precision settings 
was possible for this version of the CLASS code.

\bibliographystyle{jhep}
\bibliography{cosmobibtex}

\end{document}